\documentclass[aps,prd,eqsecnum,showpacs,superscriptaddress,twocolumn,%
nofootinbib,floatfix,preprintnumbers,altaffilletter]{revtex4}
\usepackage{graphicx}
\usepackage{hyperref}
\usepackage{amssymb}
\usepackage{amsmath}
\usepackage{longtable}
\usepackage{rotating}
\usepackage{color}
\newcommand{\xxx}[2]{\href{http://arxiv.org/abs/#1/#2}{#1/#2}}
\newcommand{\cqgref}[3]{\textit{Class.\ Quant. Grav.}\ \textbf{#1}, {#2} (#3)}
\newcommand{\prdref}[3]{\textit{Phys.\ Rev.\ D} \textbf{#1}, {#2} (#3)}
\newcommand{\prdoldref}[3]{\textit{Phys.\ Rev.\ D} \textbf{#1}, {#2} (#3)}
\newcommand{\prlref}[3]{\textit{Phys.\ Rev.\ Lett.}\ \textbf{#1}, {#2} (#3)}
\newcommand{\abs}[1]{\left|#1\right|}
\newcommand{\un}[1]{\mathrm{\,#1}}
\newcommand{\mc}[1]{\mathcal{#1}}

\newcommand{\strainLim}{1.5\times 10^{-23}\un{Hz}^{-1/2}}
\newcommand{\hsqOmegaLim}{0.53}
\newcommand{\omegaLim}{1.02}
\newcommand{\ptEst}{0.31 + 0.30i}
\newcommand{\eBar}{0.48}
\newcommand{\numSegs}{44806}
\newcommand{\effHrs}{384.1}
\newcommand{\effHrsXARM}{181.2}
\newcommand{\effHrsYARM}{114.7}
\newcommand{\effHrsNULL}{88.2}
\newcommand{\numSegsSW}{4316}
\newcommand{\effHrsSW}{37.0}
\newcommand{\effHrsSWXARM}{16.7}
\newcommand{\effHrsSWYARM}{11.2}
\newcommand{\effHrsSWNULL}{9.0}
\newcommand{\HWinj}{8100}
\newcommand{\ptEstHW}{7448 + 65i}
\newcommand{\eBarHW}{47}

\DeclareMathOperator{\Real}{Re}

\begin{document}

\title{
  First Cross-Correlation Analysis of Interferometric and Resonant-Bar\\
  Gravitational-Wave Data for Stochastic Backgrounds
}

%

\newcommand*{\ALL}{\altaffiliation[Member of ]{ALLEGRO Collaboration}} 
\newcommand*{\ALS}{\altaffiliation[Member of ]{LIGO Scientific Collaboration and ALLEGRO Collaboration}} 

\newcommand*{\AG}{Albert-Einstein-Institut, Max-Planck-Institut f\"ur Gravitationsphysik, D-14476 Golm, Germany}                                      \affiliation{\AG}                                 
\newcommand*{\AH}{Albert-Einstein-Institut, Max-Planck-Institut f\"ur Gravitationsphysik, D-30167 Hannover, Germany}                                  \affiliation{\AH}                                 
\newcommand*{\AU}{Andrews University, Berrien Springs, MI 49104 USA}                                                                                  \affiliation{\AU}                                 
\newcommand*{\AN}{Australian National University, Canberra, 0200, Australia}                                                                          \affiliation{\AN}                                 
\newcommand*{\CH}{California Institute of Technology, Pasadena, CA  91125, USA}                                                                       \affiliation{\CH}                                 
\newcommand*{\CA}{Caltech-CaRT, Pasadena, CA  91125, USA}                                                                                             \affiliation{\CA}                                 
\newcommand*{\CU}{Cardiff University, Cardiff, CF2 3YB, United Kingdom}                                                                               \affiliation{\CU}                                 
\newcommand*{\CL}{Carleton College, Northfield, MN  55057, USA}                                                                                       \affiliation{\CL}                                 
\newcommand*{\CS}{Charles Sturt University, Wagga Wagga, NSW 2678, Australia}                                                                         \affiliation{\CS}                                 
\newcommand*{\CO}{Columbia University, New York, NY  10027, USA}                                                                                      \affiliation{\CO}                                 
\newcommand*{\ER}{Embry-Riddle Aeronautical University, Prescott, AZ   86301 USA}                                                                     \affiliation{\ER}                                 
\newcommand*{\HC}{Hobart and William Smith Colleges, Geneva, NY  14456, USA}                                                                          \affiliation{\HC}                                 
\newcommand*{\IU}{Inter-University Centre for Astronomy  and Astrophysics, Pune - 411007, India}                                                      \affiliation{\IU}                                 
\newcommand*{\CT}{LIGO - California Institute of Technology, Pasadena, CA  91125, USA}                                                                \affiliation{\CT}                                 
\newcommand*{\LM}{LIGO - Massachusetts Institute of Technology, Cambridge, MA 02139, USA}                                                             \affiliation{\LM}                                 
\newcommand*{\LO}{LIGO Hanford Observatory, Richland, WA  99352, USA}                                                                                 \affiliation{\LO}                                 
\newcommand*{\LV}{LIGO Livingston Observatory, Livingston, LA  70754, USA}                                                                            \affiliation{\LV}                                 
\newcommand*{\LU}{Louisiana State University, Baton Rouge, LA  70803, USA}                                                                            \affiliation{\LU}                                 
\newcommand*{\LE}{Louisiana Tech University, Ruston, LA  71272, USA}                                                                                  \affiliation{\LE}                                 
\newcommand*{\LL}{Loyola University, New Orleans, LA 70118, USA}                                                                                      \affiliation{\LL}                                 
\newcommand*{\MS}{Moscow State University, Moscow, 119992, Russia}                                                                                    \affiliation{\MS}                                 
\newcommand*{\ND}{NASA/Goddard Space Flight Center, Greenbelt, MD  20771, USA}                                                                        \affiliation{\ND}                                 
\newcommand*{\NA}{National Astronomical Observatory of Japan, Tokyo  181-8588, Japan}                                                                 \affiliation{\NA}                                 
\newcommand*{\NO}{Northwestern University, Evanston, IL  60208, USA}                                                                                  \affiliation{\NO}                                 
\newcommand*{\RI}{Rochester Institute of Technology, Rochester, NY 14623, USA}                                                                        \affiliation{\RI}                                 
\newcommand*{\RA}{Rutherford Appleton Laboratory, Chilton, Didcot, Oxon OX11 0QX United Kingdom}                                                      \affiliation{\RA}                                 
\newcommand*{\SJ}{San Jose State University, San Jose, CA 95192, USA}                                                                                 \affiliation{\SJ}                                 
\newcommand*{\SE}{Southeastern Louisiana University, Hammond, LA  70402, USA}                                                                         \affiliation{\SE}                                 
\newcommand*{\SO}{Southern University and A\&M College, Baton Rouge, LA  70813, USA}                                                                  \affiliation{\SO}                                 
\newcommand*{\SA}{Stanford University, Stanford, CA  94305, USA}                                                                                      \affiliation{\SA}                                 
\newcommand*{\SR}{Syracuse University, Syracuse, NY  13244, USA}                                                                                      \affiliation{\SR}                                 
\newcommand*{\PU}{The Pennsylvania State University, University Park, PA  16802, USA}                                                                 \affiliation{\PU}                                 
\newcommand*{\TC}{The University of Texas at Brownsville and Texas Southmost College, Brownsville, TX  78520, USA}                                    \affiliation{\TC}                                 
\newcommand*{\TR}{Trinity University, San Antonio, TX  78212, USA}                                                                                    \affiliation{\TR}                                 
\newcommand*{\HU}{Universit{\"a}t Hannover, D-30167 Hannover, Germany}                                                                                \affiliation{\HU}                                 
\newcommand*{\BB}{Universitat de les Illes Balears, E-07122 Palma de Mallorca, Spain}                                                                 \affiliation{\BB}                                 
\newcommand*{\UA}{University of Adelaide, Adelaide, SA 5005, Australia}                                                                               \affiliation{\UA}                                 
\newcommand*{\BR}{University of Birmingham, Birmingham, B15 2TT, United Kingdom}                                                                      \affiliation{\BR}                                 
\newcommand*{\FA}{University of Florida, Gainesville, FL  32611, USA}                                                                                 \affiliation{\FA}                                 
\newcommand*{\GU}{University of Glasgow, Glasgow, G12 8QQ, United Kingdom}                                                                            \affiliation{\GU}                                 
\newcommand*{\MD}{University of Maryland, College Park, MD 20742 USA}                                                                                 \affiliation{\MD}                                 
\newcommand*{\MU}{University of Michigan, Ann Arbor, MI  48109, USA}                                                                                  \affiliation{\MU}                                 
\newcommand*{\OU}{University of Oregon, Eugene, OR  97403, USA}                                                                                       \affiliation{\OU}                                 
\newcommand*{\RO}{University of Rochester, Rochester, NY  14627, USA}                                                                                 \affiliation{\RO}                                 
\newcommand*{\SL}{University of Salerno, 84084 Fisciano (Salerno), Italy}                                                                             \affiliation{\SL}                                 
\newcommand*{\SN}{University of Sannio at Benevento, I-82100 Benevento, Italy}                                                                        \affiliation{\SN}                                 
\newcommand*{\US}{University of Southampton, Southampton, SO17 1BJ, United Kingdom}                                                                   \affiliation{\US}                                 
\newcommand*{\SC}{University of Strathclyde, Glasgow, G1 1XQ, United Kingdom}                                                                         \affiliation{\SC}                                 
\newcommand*{\WS}{University of Washington, Seattle, WA, 98195}                                                                                       \affiliation{\WS}                                 
\newcommand*{\WA}{University of Western Australia, Crawley, WA 6009, Australia}                                                                       \affiliation{\WA}                                 
\newcommand*{\UW}{University of Wisconsin-Milwaukee, Milwaukee, WI  53201, USA}                                                                       \affiliation{\UW}                                 
\newcommand*{\WU}{Washington State University, Pullman, WA 99164, USA}                                                                                \affiliation{\WU}                                 
\author{B.~Abbott}    \affiliation{\CT}                                                                                                                                                                 
\author{R.~Abbott}    \affiliation{\CT}                                                                                                                                                                 
\author{R.~Adhikari}    \affiliation{\CT}                                                                                                                                                               
\author{J.~Agresti}    \affiliation{\CT}                                                                                                                                                                
\author{P.~Ajith}    \affiliation{\AH}                                                                                                                                                                  
\author{B.~Allen}    \affiliation{\AH}  \affiliation{\UW}                                                                                                                                               
\author{R.~Amin}    \affiliation{\LU}                                                                                                                                                                   
\author{S.~B.~Anderson}    \affiliation{\CT}                                                                                                                                                            
\author{W.~G.~Anderson}    \affiliation{\UW}                                                                                                                                                            
\author{M.~Arain}    \affiliation{\FA}                                                                                                                                                                  
% ...

 \collaboration{The LIGO Scientific Collaboration [http://www.ligo.org/]
                and ALLEGRO Collaboration}
 \noaffiliation

\date{2007 March 9}

\begin{abstract}
  Data from the LIGO Livingston interferometer and the ALLEGRO
  resonant bar detector, taken during LIGO's fourth science run, were
  examined for cross-correlations indicative of a stochastic
  gravitational-wave background in the frequency range 850-950\,Hz,
  with most of the sensitivity arising between 905\,Hz and 925\,Hz.
  ALLEGRO was operated in three different orientations during the
  experiment to modulate the relative sign of gravitational-wave and
  environmental correlations.  No statistically significant
  correlations were seen in any of the orientations, and the results
  were used to set a Bayesian 90\% confidence level upper limit of
  $\Omega_{\text{gw}}(f)\le\omegaLim$, which corresponds to a
  gravitational wave strain at 915\,Hz of $\strainLim$.  In the
  traditional units of $h_{100}^2\Omega_{\text{gw}}(f)$, this is a limit of
  $\hsqOmegaLim$, two orders of magnitude better than the previous direct
  limit at these frequencies.  The method was also validated with
  successful extraction of simulated signals injected in hardware and
  software.
\end{abstract}

\preprint{LIGO-P050020-08-Z}

\maketitle

\section{INTRODUCTION}
\label{s:intro}

One of the signals targeted by the current generation of ground-based
gravitational wave (GW) detectors is a stochastic gravitational-wave
background (SGWB) \cite{Christensen:1992,Allen:1997,Maggiore:2000}.
Such a background is analogous to the cosmic microwave background,
although the dominant contribution is unlikely to have a blackbody
spectrum.  A SGWB can be characterized as cosmological or
astrophysical in origin.  Cosmological backgrounds can arise from,
for example,
pre-big-bang models
\cite{Gasperini:1993,Gasperini:2003,Buonanno:1997}, amplification of
quantum vacuum fluctuations during inflation
\cite{Grishchuk:1975,Grishchuk:1997,Starobinsky:1979}, phase
transitions \cite{Kosowsky:1992,Apreda:2002}, and cosmic strings
\cite{Caldwell:1992,Damour:2000,Damour:2005}.  Astrophysical
backgrounds consist of a superposition of unresolved sources, which
can include rotating neutron stars \cite{Regimbau:2001,Regimbau:2006},
supernovae \cite{Coward:2002} and low-mass X-ray binaries
\cite{Cooray:2004}.

\begin{figure}[htbp]
  \centering
  \includegraphics[width=240pt]{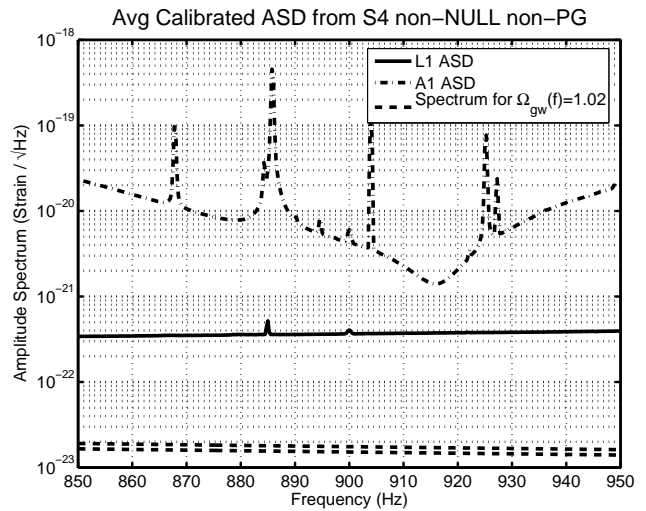}
  \caption{Sensitivity of the LLO IFO (L1) and the
    ALLEGRO bar (A1) during S4, along with strain
    associated with $\Omega_{\text{gw}}(f)=\omegaLim$ (assuming a Hubble
    constant of $H_0=72\un{km/s/Mpc}$).  (There are two
    $\Omega_{\text{gw}}(f)=\omegaLim$ curves,
    corresponding to the different strain levels
    such a background would generate
    in an IFO and a bar, as explained in Sec.\ref{s:sgwb}
    and \protect\cite{Whelan:2006}.)
    The quantity
    plotted is amplitude spectral density (ASD), the square root of
    the one-sided power spectral density defined in
    \protect\eqref{e:s1s1corr}, at a resolution of 0.25\,Hz.}
\label{fig:avg_psds}
\end{figure}

The standard cross-correlation search \cite{Allen:1999} for a
SGWB necessarily requires two or more GW detectors.  Such searches
have been performed using two resonant bar detectors \cite{Astone:1999}
and also using two or more kilometer-scale GW interferometers
(IFOs) \cite{s1sgwb,s3sgwb,s4sgwb}.  The present work describes the results of
the first cross-correlation analysis carried out between an IFO (the
4\,km IFO at the
LIGO Livingston Observatory (LLO),
known as L1) and a bar (the cryogenic
ALLEGRO detector, referred to as A1).
This pair of detectors is separated by only
40\,km, the closest pair among modern ground-based GW detector
sites, which allows it to probe the stochastic GW spectrum around
900\,Hz.  In addition, the ALLEGRO bar can be rotated, changing the
response of the correlated data streams to stochastic GWs and thus
providing a means to distinguish correlations due to a SGWB from those
due to correlated environmental noise \cite{Finn:2001}.
This paper describes cross-correlation analysis of L1 and A1 data 
taken between February 22 and March 23, 2005, during
LIGO's fourth science run (S4).  Average sensitivities of L1 and
A1 during S4 are shown in Fig.~\ref{fig:avg_psds}.
ALLEGRO was operated in three
orientations, which modulated the GW response of the LLO-ALLEGRO pair
through $180^\circ$ of phase.

The LLO-ALLEGRO correlation experiment is complementary to experiments
using data from the two LIGO sites, in that it is sensitive to a SGWB
at frequencies of around 900\,Hz rather than 100\,Hz.  Targeted
sources are thus those with a relatively narrow-band spectrum peaked
near 900\,Hz.  Spectra with such shapes can arise from exotic
cosmological models, as described in Sec.~\ref{s:sgwb}, or from
astrophysical populations \cite{Regimbau:2006}.

The organization of this paper is as follows.  Sec.~\ref{s:sgwb}
reviews the properties and characterization of a SGWB.
Sec.~\ref{s:expt} describes the LLO and ALLEGRO experimental
arrangements, including the data acquisition and strain calibration
for each instrument.  Sec.~\ref{s:method} describes the
cross-correlation method and its application to the present situation.
Sec.~\ref{s:post} describes the
details of the post-processing methods and statistical interpretation
of the cross-correlation results.  Sec.~\ref{s:res} describes the
results of the cross-correlation measurement and the corresponding
upper limit on the SGWB strength in the range 850--950\,Hz.
Sec.~\ref{s:inj} describes the results of our analysis pipeline when
applied to simulated signals injected both within the analysis
software and in the hardware of the instruments themselves.
Sec.~\ref{s:others} compares our results to those of previous
experiments and to the sensitivities of other operating detector
pairs.
Sec.~\ref{s:future} considers the prospects for future work.

\section{STOCHASTIC GRAVITATIONAL-WAVE BACKGROUNDS}
\label{s:sgwb}

A gravitational wave (GW) is described by the metric tensor
perturbation $h_{ab}(\vec{r},t)$.  A given GW detector, located at
position $\vec{r}_{\text{det}}$ on the Earth, will measure a GW strain
which, in the long-wavelength limit, is some projection of this tensor:
\begin{equation}
  \label{e:straindfn}
  h(t) = h_{ab}(\vec{r}_{\text{det}},t) d^{ab}
\end{equation}
where $d^{ab}$ is the detector response tensor, which is
\begin{equation}
  \label{e:difo}
  d_{\text{(ifo)}}^{ab} = \frac{1}{2}(\hat{x}^a \hat{x}^b - \hat{y}^a \hat{y}^b)
\end{equation}
for an interferometer with arms parallel to the unit vectors $\hat{x}$
and $\hat{y}$ and
\begin{equation}
  \label{e:dbar}
  d_{\text{(bar)}}^{ab} = \hat{u}^a \hat{u}^b
\end{equation}
for a resonant bar with long axis parallel to the unit vector
$\hat{u}$.

\begin{figure}[htbp]
  \centering
  \includegraphics[width=240pt]{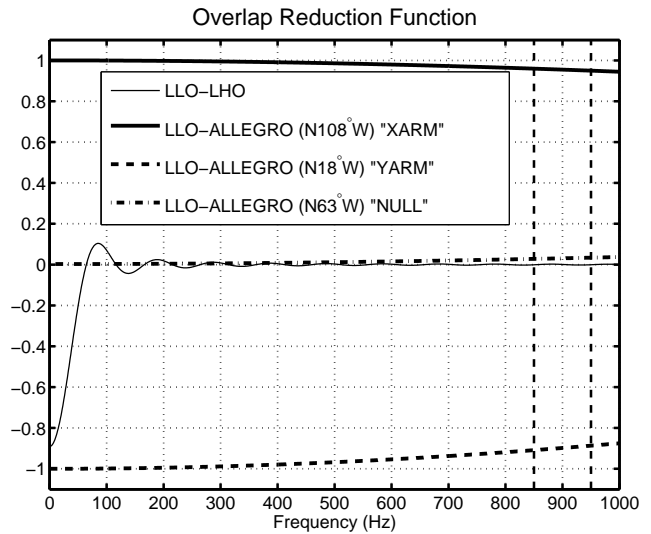}
  \caption{The overlap reduction function for LIGO Livingston
    Observatory (LLO) with ALLEGRO and with LIGO Hanford Observatory
    (LHO).  The three LLO-ALLEGRO curves correspond to the three
    orientations in which ALLEGRO was operated during LIGO's S4 run:
    ``XARM'' (N$108^\circ$W) is nearly parallel to the x-arm of LLO
    (``aligned''); ``YARM'' (N$18^\circ$W) is nearly parallel to the
    y-arm of LLO (``anti-aligned''); ``NULL'' (N$63^\circ$W) is
    halfway in between these two orientations (a ``null alignment''
    midway between the two LLO arms).
    Note that for non-zero frequencies, the separation vector between
    the two sites breaks the symmetry between the ``XARM'' and
    ``YARM'' alignments, and leads to an offset of the ``NULL''
    curve, as described in \protect\cite{Whelan:2006}.
    The LLO-LHO overlap reduction
    function is shown for reference.  The frequency band of the
    present analysis, $850\un{Hz}\le f\le 950\un{Hz}$, is indicated
    with dashed vertical lines.}
\label{fig:overlap}
\end{figure}

A stochastic GW background (SGWB) can arise from a superposition of
uncorrelated cosmological or astrophysical sources.  Such a
background, which we assume to be isotropic, unpolarized, stationary and Gaussian, will
generate a cross-correlation between the strains measured by two
detectors.  In terms of the continuous Fourier transform defined by
$\widetilde{a}(f)=\int_{-\infty}^\infty dt\,a(t)\,\exp(-i2\pi ft)$, the expected
cross-correlation is
\begin{equation}
  \label{e:h1h2corr}
  \langle \widetilde{h}_1^*(f) \widetilde{h}_2(f') \rangle
  = \frac{1}{2} \delta(f-f')\,S_{\text{gw}}(f)\,\gamma_{12}(f)
\end{equation}
where
\begin{equation}
  \gamma_{12}(f)={d_{1ab}}\, {d_2^{cd}}\,
  \frac{5}{4\pi} \iint d^2\Omega_{\hat{n}}\,
  P^{\text{TT}\hat{n}}{}^{ab}_{cd}\,
  e^{i2\pi f\hat{n}\cdot(\vec{r}_2-\vec{r}_1)/c}
\end{equation}
is the overlap reduction function (ORF) \cite{Flanagan:1993}
between the two
detectors, defined in terms of the projector
$P^{\text{TT}\hat{n}}{}^{ab}_{cd}$ onto traceless symmetric tensors
transverse to the unit vector $\hat{n}$.  The ORF for several detector
pairs of interest is shown in Fig.~\ref{fig:overlap}.

$S_{\text{gw}}(f)$ is the one-sided spectrum of the SGWB.  This is the
one-sided power spectral density (PSD) the background would generate
in an interferometer with perpendicular arms, which can be seen from
\eqref{e:h1h2corr} and the fact that the ORF of
such an interferometer with itself is unity.  Since the ORF
of a resonant bar with itself is $4/3$ (see
\cite{Whelan:2006} and Sec.~\ref{ss:inj-alg} for more details), the PSD of
the strain measured by a bar detector due to the SGWB would be
$(4/3)S_{\text{gw}}(f)$.

A related measure of the spectrum is the dimensionless quantity
$\Omega_{\text{gw}}(f)$, the GW energy density per unit logarithmic
frequency divided by the critical energy density $\rho_{\text{c}}$
needed to close the universe:
\begin{equation}
  \label{e:Omega_gw}
  \Omega_{\text{gw}}(f)= \frac{f}{\rho_{\text{c}}}\,
  \frac{d\rho_{\text{gw}}}{df}
  = \frac{10\pi^2}{3 H_0^2}\,f^{3} S_{\text{gw}}(f)
  \ .
\end{equation}
Note that the definition $\Omega_{\text{gw}}(f)$ thus depends on the
value of the Hubble constant $H_0$.  Most SGWB literature avoids this
artificial uncertainty by working in terms of
\begin{equation}
  h_{100}^2\Omega_{\text{gw}}(f) =
  \left(
    \frac{H_0}{100\un{km/s/Mpc}}
  \right)^2
  \Omega_{\text{gw}}(f)
\end{equation}
rather than $\Omega_{\text{gw}}(f)$ itself.  We will instead follow
the precedent set by \cite{s3sgwb} and quote numerical values for
$\Omega_{\text{gw}}(f)$ assuming a Hubble constant of
$72\un{km/s/Mpc}$.

A variety of spectral shapes have been proposed for $\Omega_{\text{gw}}$, 
for both astrophysical and cosmological
stochastic backgrounds \cite{Maggiore:2000,coward,ferrari}. For
example, whereas the slow-roll inflationary model predicts a constant
$\Omega_{\text{gw}}(f)$ in the bands of LIGO or ALLEGRO, certain
alternative cosmological models predict broken-power law spectra,
where the rising and falling slopes, and the peak-frequency are
determined by model parameters \cite{Maggiore:2000}.
String-inspired pre-big-bang cosmological models belong to this category
\cite{Gasperini:2003,Brustein:1995ah}.
For certain ranges of these three parameters the LLO-ALLEGRO
correlation measurement offers the best constraints on theory that can
be inferred from any contemporary observation.
This can happen, e.g., if the 
power-law exponent on the rising spectral
slope is greater than 3 and the peak-frequency is sufficiently close
to 900 Hz \cite{Bose:2005fm}.

\section{EXPERIMENTAL SETUP}
\label{s:expt}

\subsection{The LIGO Livingston Interferometer}
\label{ss:expt-LLO}

The experimental setup of the LIGO observatories has been described at
length elsewhere \cite{s1ligo}. Here we provide a brief review, with
particular attention paid to details significant for the LLO-ALLEGRO
cross-correlation measurement.

The LIGO Livingston Observatory (LLO) is an
interferometric GW detector with perpendicular 4-km arms.
The laser interferometer senses directly any changes in the
differential arm length.  It does this by splitting a light beam at
the vertex, sending the separate beams into 4-km long optical cavities of their
respective arms, and then recombining the beams to detect any
change in the optical phase difference between the arms, which is equivalent to a
difference in light travel time.
This provides a measurement of $h(t)$ as defined in
\eqref{e:straindfn} and \eqref{e:difo}.  However, the measured
quantity is not exactly $h(t)$ for two reasons:

First, there are local forces which perturb the test masses, and so
produce changes in arm length.  There are also optical and
electronic fluctuations that mimic real strains.
The combination of these effects causes a strain noise $n(t)$ to
always be present in the output, producing a measurement of
\begin{equation}
s(t) = h(t) + n(t)
\end{equation}

Second, the test masses are not really
free.  There is a servo system, which uses changes in the differential
arm length as its error signal $q(t)$, and then applies extra
(``control'') forces to the test masses to keep the
differential arm length nearly zero.  It is this error signal $q(t)$
which is recorded, and its relationship to the strain
estimate $s(t)$ is most easily described in the Fourier domain:
\begin{equation}
  \widetilde{s}(f) = \widetilde{R}(f) \widetilde{q}(f)
\end{equation}
The response function $\widetilde{R}(f)$
is estimated by a combination of modelling and
measurement \cite{s4cal} and varies slowly over the course of the
experiment.

Because the error signal $q(t)$ has a smaller dynamic range than the
reconstructed strain $s(t)$, our analysis starts from the digitized
time series $q[k]=q(t_k)$ (sampled $2^{14} = 16384$ times per second,
and digitally downsampled to $4096\un{Hz}$ in the analysis) and
reconstructs the LLO strain only in the frequency domain.

\subsection{The ALLEGRO Resonant Bar Detector}
\label{ss:expt-ALLEGRO}

The ALLEGRO resonant detector, operated by a group from Louisiana State 
University \cite{ALLEGRO},  is a two-ton aluminum cylinder coupled to a 
niobium secondary resonator. The secondary resonator is part of an inductive 
transducer \cite{Harry:1999} which is coupled to a DC SQUID.  Strain along the cylindrical
axis excites the first longitudinal vibrational mode of the bar. The transducer
is tuned for sensitivity to this mechanical mode. Raw data 
acquired from the detector thus reflect the high $Q$ resonant mechanical 
response of the system. A major technical challenge of this analysis is
due to the extent to which the bar data differ from those of the interferometer.

\subsubsection{Data Acquisition, Heterodyning and Sampling}
\label{sss:expt-ALLEGRO-het}

The ALLEGRO detector has a relatively narrow sensitive band of
$\sim$~100\,Hz centered around $\sim$~900\,Hz near the two normal modes of the
mechanical bar-resonator
system. For this reason, the output of the detector can first be
heterodyned with a commercial lock-in amplifier
to greatly reduce the sampling rate, which is set at
250~samples/s. Both the in-phase and quadrature outputs of the lock-in
are recorded and the detector output can thus be represented as
a complex time series which covers a 250\,Hz band centered on the
lock-in reference oscillator frequency. This reference frequency is
chosen to be near the center of the sensitive band, and during the S4
run it was set to 904\,Hz. The overall timing of data heterodyned 
in this fashion is provided by both the sampling clock and the reference oscillator. 
Both time bases were locked to GPS. 

The nature of the resonant detector and its data acquisition system gives
rise to a number of timing
issues: heterodyning, filter delays of the electronics and
the timing of the data acquisition system itself \cite{s4ALLEGROcal}.
It is of critical importance that the timing be fully understood
so that the phase of any potential signal may be recovered.
Convincing evidence that all of the issues are accounted for in
is demonstrated by the recovery and
cross-correlation of test signals simultaneously injected into both
detectors. The signals were recovered at the expected phase as
presented in Sec.~\ref{s:inj}.

\subsubsection{Strain Calibration}
\label{sss:expt-ALLEGRO-cal}
\begin{figure}[htbp]
  \centering
  \includegraphics[width=240pt]{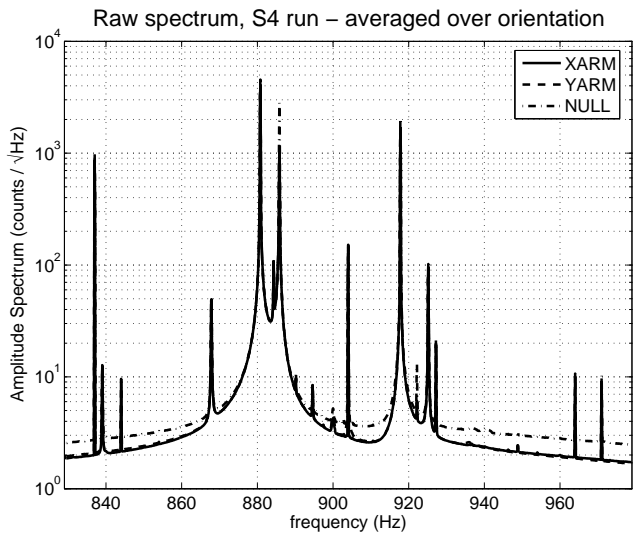}
\caption{The graph displays the amplitude spectral density of raw ALLEGRO 
detector output during S4, at a frequency resolution of 0.1\,Hz.
For this graph these data have not been 
transformed to strain via the calibrated response function 
The vertical scale represents digital counts/$\sqrt{\mbox{Hz}}$
The normal mechanical modes where
the detector is most sensitive are at 880.78\,Hz and 917.81\,Hz. There is an injected calibration
line at 837\,Hz. Also prominent are an extra mechanical resonance at 885.8\,Hz and a peak at 904\,Hz (DC in the heterodyned data stream)}
\label{fig:ALrawnoise}
\end{figure}

The raw detector output is proportional to the displacement of the secondary resonator,  and thus has a spectrum with sharp line
features due to the high-Q resonances of the bar-resonator system
as can be seen in Fig.~\ref{fig:ALrawnoise}. 
The desired GW
signal is the effective strain on the bar, and recovering this means
undoing the resonant response of the detector. This response has
a long coherence time -- thus long stretches of data are needed to resolve
the narrow lines in the raw data. The strain data
have a much flatter spectrum,
as shown in Fig.~\ref{fig:avg_psds}. Therefore it is practical to generate 
the calibrated strain time series,
$s(t)$, and use that as the input to the cross correlation
analysis. 

The calibration procedure, described in detail in
\cite{s4ALLEGROcal}, is carried out in the frequency domain and
consists of the following: A 30 minute stretch of clean ALLEGRO data is
windowed and Fourier-transformed.
The mechanical mode frequencies drift slightly due
to small temperature variations, so these frequencies are determined
for each stretch and those are incorporated into the model of the
mechanical response of the system to a strain. The model consists
of two double poles
at these normal mode frequencies. In addition to this response
we must then account for the phase shifts due to the time delays in the
lock-in and anti-aliasing filters. 

After applying the full response function, the data are then inverse
Fourier-transformed back to the time domain.
The next 50\% overlapping 30 minute
segment is then taken. The windowed segments are stitched together until
the entire continuous stretch of good data is completed. The first and
last 15 minutes are dropped. The result represents a heterodyned
complex time series of strain, whose amplitude spectral density is
shown in Fig.~\ref{fig:avg_psds}.

The overall scale of the detector output in terms of strain is 
determined by applying a
known signal to the bar. A force applied to one end of the bar has a simple theoretical relationship
to an equivalent gravitational strain
\cite{s4ALLEGROcal,McHugh:2005,Morse:1997}.
A calibrated force can be applied via a capacitive
``force generator'' which also provides the mechanism used for hardware signal
injections.  A reciprocal measurement -- excitation followed by 
measurement with the same transducer -- along with known properties 
of the mechanical system, allows the determination of the force 
generator constant . With that constant determined  (with units of newtons per volt) a calibrated force is applied to the bar and the overall scale of the response determined.

\begin{table*}
    \begin{tabular}{||c||c||c||c|c|c||}
    \hline
    \hline
Dates &  orientation &  azimuth & $\gamma(850\un{Hz})$
& $\gamma(915\un{Hz})$ & $\gamma(950\un{Hz})$\\
    \hline
    \hline
2005 Feb 22--2005 Mar  4 &      YARM & N$108^\circ$W & {-0.9087} & {-0.8947} & {-0.8867}\\
    \hline
2005 Mar  4--2005 Mar 18 &      XARM & N$18^\circ$W & {0.9596} & {0.9533} & {0.9498}\\
    \hline
2005 Mar 18--2005 Mar 23 &      NULL & N$63^\circ$W & {0.0280} & {0.0318} & {0.0340}\\
    \hline
    \hline
  \end{tabular}

  \caption{
    Orientations of ALLEGRO during the LIGO S4 Run, including overlap
    reduction function evaluated at the extremes of the analyzed
    frequency range, and at the frequency of peak sensitivity.
    Note that while the NULL orientation represents perfect
    misalignment ($\gamma=0$) at $0\un{Hz}$, it is not quite perfect at
    the frequencies of interest.  This is primarily because of an
    azimuth-independent offset term in $\gamma(f)$ which contributes
    at non-zero frequencies \cite{Whelan:2006,Finn:2001}. Due to this
    term, it is impossible to orient ALLEGRO so that $\gamma(f)=0$ at
    all frequencies, and to set it to zero around $915\un{Hz}$ one would
    have to use an azimuth of N$62^\circ$W rather than N$63^\circ$W.
  }
  \label{tab:orientations}
\end{table*}

\subsubsection{Orientation}

A unique feature of this experiment is the ability to rotate the ALLEGRO
detector and modulate the response of the ALLEGRO-LLO pair to a GW
background \cite{Finn:2001}.
Data were taken in three different orientations of ALLEGRO, known as
XARM, YARM, and NULL, detailed in Table~\ref{tab:orientations}.
As shown in Fig.~\ref{fig:overlap} and
\eqref{e:h1h2corr}, these orientations correspond to different
pair responses due to different overlap reduction functions.  In the
XARM orientation--the bar axis parallel to the x-arm of the
interferometer--a GW signal produces positive correlation between the
data in the two detectors
detectors. In the YARM orientation a GW signal produces an
anti-correlation. In the NULL orientation--the bar aligned halfway
between the two arms of the interferometer--the pair has very nearly
zero sensitivity as a GW signal produces almost zero correlation
between the detectors' data. A real signal is thus modulated whereas many
types of instrumental correlation would not have the same dependence
on orientation.

\section{CROSS-CORRELATION METHOD}
\label{s:method}

This section describes the method to used to search for a SGWB by
cross-correlating detector outputs.  In the case of L1-A1 correlation
measurements, it is complicated by the different sampling rates for
the two instruments and the fact that the A1 data are heterodyned at
904\,Hz prior to digitization.

\subsection{Continuous-Time Idealization}

Both ground-based interferometric and resonant-mass detectors produce
a time-series output which can be related to a discrete sampling of
the signal
\begin{equation}
  s_i(t) = h_i(t) + n_i(t)
\end{equation}
where $i$ labels the detector (1 or 2 in this case), $h_i(t)$ is the
gravitational-wave strain defined in \eqref{e:straindfn}, and $n_i(t)$
is the instrumental noise associated with each detector, converted
into an equivalent strain.  The detector output is characterized by
its power spectral density $P_i(f)$
\begin{equation}
  \label{e:s1s1corr}
  \langle \widetilde{s}_i^*(f) \widetilde{s}_i(f')\rangle
  = \frac{1}{2} \delta(f-f')\,P_i(f)
\end{equation}
which should be dominated by the auto-correlation of the noise
($\langle \widetilde{s}_i^*(f) \widetilde{s}_i(f')\rangle \approx
\langle \widetilde{n}_i^*(f) \widetilde{n}_i(f')\rangle$).  If the
instrument noise is approximately uncorrelated, the expected
cross-correlation of the detector outputs is [\textit{cf.}\
\eqref{e:h1h2corr}]
\begin{equation}
  \label{e:s1s2corr}
  \begin{split}
  \langle \widetilde{s}_1^*(f) \widetilde{s}_2(f')\rangle
  &
  \approx \langle \widetilde{h}_1^*(f) \widetilde{h}_2(f')\rangle
  \\
  &
  = \frac{1}{2} \delta(f-f')\,S_{\text{gw}}(f)\,\gamma_{12}(f)    
  \end{split}
\end{equation}
which can be used along with the auto-correlation \eqref{e:s1s1corr}
to determine the statistical properties of the cross-correlation
statistic defined below.

We use the optimally-filtered cross-correlation method described in
\cite{s1sgwb,Allen:1999} to calculate a cross-correlation
statistic which is an approximation to the continuous-time
cross-correlation statistic
\begin{equation}
  \begin{split}
  Y^c
  &=
  \int dt_1\, dt_2\, {s_1(t_1)}\, {Q(t_1-t_2)}
  {s_2(t_2)}
  \label{e:CCstat}
  \\
  &=
  \int df\,{\widetilde{s}_1^*(f)}\, {\widetilde{Q}(f)}\,
  {\widetilde{s}_2(f)}
  \end{split}
\end{equation}
In the continuous-time idealization, such a cross-correlation
statistic, calculated over a time $T$, has an expected mean
\begin{equation}
  \label{e:contmean}
  \mu_{Y^c} = \langle Y^c\rangle 
  \approx \frac{T}{2}
  \int_{-\infty}^{\infty} df \,\gamma(\abs{f})\,S_{\text{gw}}(f)\widetilde{Q}(f)
\end{equation}
and variance
\begin{equation}
  \label{e:contvar}
  \sigma_{Y^c}^2 = \langle (Y^c-\mu_{Y^c})^2\rangle
  \approx \frac{T}{4}
  \int_{-\infty}^{\infty} df \,P_{1}(f)\,P_{2}(f)\abs{\widetilde{Q}(f)}^2
\end{equation}
Using \eqref{e:contmean} and \eqref{e:contvar}, the optimal choice
for the filter $\widetilde{Q}(f)$, given a predicted shape for the
spectrum $S_{\text{gw}}(f)$ can be shown \cite{Allen:1999} to be
\begin{equation}
  \label{e:Qopt}
  \widetilde{Q}(f) \propto \frac{\gamma(\abs{f})\,S_{\text{gw}}(f)}
  {P_{1}(f)\,P_{2}(f)}
\end{equation}
If the spectrum of gravitational waves is assumed to be a power law
over the frequency band of interest, a convenient parameterization of
the spectrum, in terms of $\Omega_{\text{gw}}(f)$ defined in
\eqref{e:Omega_gw}, is
\begin{equation}
  \Omega_{\text{gw}}(f) = \Omega_R \left(\frac{f}{f_R}\right)^\alpha
\end{equation}
where $f_R$ us a conveniently-chosen reference frequency and
$\Omega_R=\Omega_{\text{gw}}(f_R)$.
The cross-correlation measurement is then a measurement of $\Omega_R$,
and if the optimal filter is normalized according to
\begin{subequations}
\label{eq:Qoptnorm}
\begin{equation}
  \widetilde{Q}(f) = \mc{N}\,
  \frac{\gamma(f)\,(f/f_R)^\alpha}{|f|^3\,P_1(f)\,P_2(f)}
\end{equation}
where
\begin{equation}
  \mc{N} = \frac{20\pi^2}{3\,{H_0}^2}
  \left(
    \int_{-\infty}^\infty \frac{df}{f^6} 
    \frac{[\gamma(f)\,(f/f_R)^\alpha]^2}
    {P_1(f)P_2(f)}
  \right)^{-1}
\end{equation}
\end{subequations}
then the expected statistics of $Y^c$ in the presence of a background
of actual strength $\Omega_R$ are
\begin{equation}
  \mu_{Y^c} = \Omega_R\,T
\end{equation}
and
\begin{equation}
  \label{e:sigmatheor}
  \sigma_{Y^c}^2 = T\,\left(\frac{10\pi^2}{3\,{H_0}^2}\right)^2
  \left(
    \int_{-\infty}^\infty \frac{df}{f^6}
    \frac{[\gamma(f)\,(f/f_R)^\alpha]^2}
    {P_1(f)P_2(f)}
  \right)^{-1}
\end{equation}
and a measurement of $Y^c/T$ provides a \textit{point estimate} of the
background strength $\Omega_R$ with associated estimated errorbar
$\sigma_{Y^c}/T$.

\subsection{Discrete-Time Method}
\label{ss:method-disc}

\subsubsection{Handling of Different Sampling Rates and Heterodyning}
\label{sss:method-disc-het}

Stochastic-background measurements using pairs of LIGO
interferometers \cite{s1sgwb} have implemented \eqref{e:CCstat}
from discrete samplings $s_i[k]=s(t_0+k\,\delta t)$ as follows: First
the continuous Fourier transforms $\widetilde{s}(f)$ were approximated using
the discrete Fourier transforms of windowed and zero-padded versions
of the discrete time series; then an optimal filter was constructed
using an approximation to \eqref{e:Qopt}, and finally the product of
the three was summed bin-by-bin to approximate the integral over
frequencies.  The
discrete version of $\widetilde{Q}(f)$ was simplified in two ways:
first, because of the averaging used in calculating the power
spectrum, the frequency resolution on the optimal filter was generally
coarser than that associated with the discrete Fourier transforms of
the data streams, and second, the value of the optimal filter was
arbitrarily set to zero outside some desired range of frequencies
$f_{\text{min}}\le f\le f_{\text{max}}$.  This was justified because
the optimal filter tended to have little support for frequencies
outside that range.

The present experiment has two additional complications associated
with the discretization of the time-series data.  First, the A1
data are not a simple time-sampling of the gravitational-wave strain,
but are base-banded with a heterodyning frequency $f^h_2=904\un{Hz}$
as described in
Sec.~\ref{sss:expt-ALLEGRO-het}--\ref{sss:expt-ALLEGRO-cal}.  Second,
the A1 data are sampled at $(\delta{t}_2)^{-1}=250\un{Hz}$, while
the L1 data are sampled at 16384\,Hz, and subsequently downsampled to
$(\delta{t}_1)^{-1}=4096\un{Hz}$.  This would make a time-domain
cross-correlation extremely problematic, as it would necessitate a
large variety of time offsets $t_1-t_2$.  In the frequency domain, it
means that the downsampled L1 data, once calibrated, represent
frequencies ranging from $-2048\un{Hz}$ to $2048\un{Hz}$, while the
calibrated A1 data represent frequencies ranging from
$(904-125)\un{Hz}=779\un{Hz}$ to $(904+125)\un{Hz}=1029\un{Hz}$.
These different frequency ranges do not pose a problem, as long as the
range of frequencies chosen for the integral satisfies
$f_{\text{min}}> 779\un{Hz}$ and $f_{\text{max}}< 1029\un{Hz}$. Another
requirement is that for the chosen frequency resolution, the A1 data 
heterodyne reference frequency must align with a frequency bin in the L1 data.
This is satisfied for integer-second data spans and integer-hertz reference frequencies.
With these conditions, the Fourier transforms of the A1 and L1 data are
both defined over a common set of frequencies, as 
detailed in \cite{Whelan:2005}.
Looking at the A1 spectrum in Fig.~\ref{fig:avg_psds}, a reasonable
range of frequencies should be a subset of the range $850\un{Hz}\lesssim
f\lesssim 950\un{Hz}$.

\subsubsection{Discrete-Time Cross-Correlation}
\label{sss:method-disc-crosscorr}

Explicitly, the time-series inputs to the analysis pipeline, from each
$T=60\un{sec}$ of analyzed data, are:
\begin{itemize}
\item For L1, a real time series $\{q_1[j]|j=0,\ldots N_1-1\}$,
  sampled at $(\delta{t}_1)^{-1}=4096\un{Hz}$, consisting of $N_1 =
  T/\delta{t}_1 = 245760$ points.  This is obtained by downsampling
  the raw data stream by a factor of 4.  The data are downsampled to
  $4096\un{Hz}$ rather than $2048\un{Hz}$ to ensure that the rolloff
  of the associated anti-aliasing filter is outside the frequency
  range being analyzed.  The raw L1 data are related to
  gravitational-wave strain by the calibration response function
  $\widetilde{R}_1(f)$ constructed as described in
  Sec.~\ref{ss:expt-LLO}.
\item For A1, a complex time series $\{s_2^h[k]|k=0,\ldots
  N_2-1\}$, sampled at $(\delta{t}_2)^{-1}=250\un{Hz}$, consisting of
  $N_2 = T/\delta{t}_2 = 15000$ points.  This is calibrated to
  represent strain data as described in
  Sec.~\ref{sss:expt-ALLEGRO-cal}, but still heterodyned.
\end{itemize}
To produce an approximation of the Fourier transform of the data
from detector $i$,
the data are multiplied by an appropriate windowing
function, zero-padded to twice their original length,
discrete-Fourier-transformed, and multiplied by
$\delta{t}_i$.
In addition, the L1 data are multiplied by a
calibration response function, while the A1 data are interpreted
as representing frequencies appropriate in light of the heterodyne.
For L1,
\begin{multline}
  \widetilde{s}_1(f_\ell) \sim \widetilde{s}_1[\ell]
  \\
  := \widetilde{R}_1(f_\ell)
  \sum_{j=0}^{N_1-1} w_1[j]q_1[j]
  \exp
  \left(
    \frac{- i\,2\pi\,\ell j}{2N_1}
  \right)
  \,\delta t_1
  \ ,
  \\
  \ell = -N_1,\ldots,N_1-1
\end{multline}
where $f_\ell=\frac{\ell}{2T}$ is the frequency associated with the
$\ell$th frequency bin.  In the case of A1, the identification is
offset by $\ell_2^h=f_h\cdot(2T)=107880$:
\begin{multline}
  \widetilde{s}_2(f_\ell) \sim \widetilde{s}_2[\ell]
  \\
  := \sum_{k=0}^{N_2-1} w_2[k]q_2[k]
  \exp
  \left(
    \frac{- i\,2\pi\,(\ell-\ell_2^h) k}{2N_2}
  \right)
  \,\delta t_2
  \ ,
  \\
  \ell = \ell_2^h -N_2,\ldots,\ell_2^h + N_2 - 1    
\end{multline}
As is shown in \cite{Whelan:2005}, if we construct a cross-correlation
statistic
\begin{equation}
  \label{eq:discfreqstat}
  Y := \sum_{\ell=\ell_{\text{min}}}^{\ell_{\text{max}}}
  \frac{1}{2T} \, [\widetilde{s}_1(f_\ell)]^*\,
  \widetilde{Q}(f_\ell)\,\widetilde{s}_2(f_\ell)
\end{equation}
the expected mean value in the presence of a stochastic background
with spectrum $S_{\text{gw}}(f)$ is
\begin{equation}
  \label{eq:mugood}
  \mu := \langle Y\rangle \approx \overline{w_1w_2}\,\frac{T}{2}\,
  \sum_{\ell=\ell_{\text{min}}}^{\ell_{\text{max}}}
  \frac{1}{2T}\,[\widetilde{Q}(f_\ell)]\,\gamma(f_\ell)\,S_{\text{gw}}(f_\ell)
\end{equation}
where $\overline{w_1w_2}$ is an average of the product of the two
windows, calculated using the points which exist at both sampling
rates; specifically, if $r_1$ and $r_2$ are the smallest integers such
that $\delta{t}_1/\delta{t}_2 = r_1/r_2 = 125/2048$,
\begin{equation}
  \overline{w_1w_2} = \frac{r_1r_2}{N}
  \sum_{n=0}^{N/(r_1r_2)-1} w_1[nr_2]w_2[nr_1]
\end{equation}
Note that while the average value given by \eqref{eq:mugood} is
manifestly real, any particular measurement of $Y$ will be complex,
because of the band-pass associated with the heterodyning of the A1
data.  As shown in \cite{Whelan:2005}, the real and imaginary parts of
the cross-correlation statistic each have expected variance
\begin{equation}
  \begin{split}
  \sigma^2
  &
  := \frac{1}{2}\langle Y^* Y\rangle
  \\
  &
  \approx
  \frac{T}{8}\,\overline{w_1^2w_2^2}
  \sum_{\ell=\ell_{\text{min}}}^{\ell_{\text{max}}}\frac{1}{2T}\,
  \abs{\widetilde{Q}(f_\ell)}^2\,
  P_1(\abs{f_\ell})\, P_2(\abs{f_\ell})    
  \end{split}
\end{equation}
where once again $\overline{w_1^2w_2^2}$ is an average over the time
samples the two windows have ``in common'':
\begin{equation}
  \overline{w_1^2w_2^2} = \frac{r_1r_2}{N}
  \sum_{n=0}^{N/(r_1r_2)-1} (w_1[nr_2])^2(w_2[nr_1])^2
\end{equation}

\subsubsection{Construction of the Optimal Filter}
\label{sss:method-disc-optimal}

To perform the cross-correlation in \eqref{eq:discfreqstat}, we need
to construct an optimal filter by a discrete approximation to
\eqref{eq:Qoptnorm}.  We approximate the power spectra $P_1(f)$ using
Welch's method \cite{Welch:1967}; as a consequence of the averaging of
periodograms constructed from shorter stretches of data, the power
spectra are estimated with a frequency resolution $\delta\!f$ which is
coarser than the $1/2T$ associated with the construction in
\eqref{eq:discfreqstat}.  As detailed in \cite{s1sgwb}, we
handle this by first multiplying together
$[\widetilde{s}_1(f_\ell)]^*$ and $\widetilde{s}_2(f_\ell)$ at the
finer frequency resolution, then summing together sets of
$2T\,\delta\!f$ bins and multiplying them by the coarser-grained
optimal filter.  For our search, $\delta\!f=0.25\un{Hz}$ and
$T=60\un{sec}$, so $2T\,\delta\!f=30$.

\subsubsection{Power Spectrum Estimation}
\label{sss:method-disc-psd}

Because the noise power spectrum of the LLO
can vary with time, we continuously update the optimal filter used in
the cross-correlation measurement.  However, using an optimal filter
constructed from power spectra calculated from the same data to be
analyzed leads to a bias in the cross-correlation statistic $Y$,
as detailed in \cite{bias}.  To avoid this,
we analyze each $T=60\un{sec}$ segment of data using an optimal filter
constructed from the average of the power spectra from the segments
before and after the segment to be analyzed.  This method is known as
``sliding power spectrum estimation'' because, as we analyze
successive segments of data, the segments used to calculated the PSDs
for the optimal filter slide through the data to remain adjacent.  The
$\delta\!f=0.25\un{Hz}$ resolution is obtained by calculating the
power spectra using Welch's method with 29 overlapped 4-second
sub-segments in each 60-second segment of data, for a total of 58
sub-segments.

\section{POST-PROCESSING TECHNIQUES}
\label{s:post}

\subsection{Stationarity Cut}
\label{ss:post-sigmarat}

The sliding power spectrum estimation method described in
Sec.~\ref{sss:method-disc-psd} can lead to inaccurate results if the
noise level of one or both instruments varies widely over successive
intervals.  Most problematically, if the data are noisy only within a
single analysis segment, consideration of the power spectrum
constructed from the segments before and after, which are not noisy,
will cause the segment to be over-weighted when combining
cross-correlation data from different segments.  To avoid this, we
calculate for each segment both the usual estimated standard
deviation $\sigma_I$ using the ``sliding'' PSD estimator and the
``na\"{\i}ve'' estimated standard deviation $\sigma_I'$ using the data
from the segment itself.  If the ratio of these two is too far from
unity, the segment is omitted from the cross-correlation analysis.
The threshold used for this analysis was 20\%.
The amount of data excluded based on this cut was between 1\% and 2\%
in each of the three orientations, and subsequent investigations show
the final results would not change significantly for any reasonable
choice of threshold.

\subsection{Bias-Correction of Estimated Errorbars}
\label{ss:post-bias}

Although use of the sliding power spectrum estimator removes any bias
from the optimally-filtered cross-correlation measurement, the methods
of \cite{bias} show that there is still a slight underestimation of
the estimated standard deviation associated with the finite number
of periodograms averaged together
in calculating the power spectrum.  To correct for
this, we have to scale up the errorbars by a factor of
$1+1/(N_{\text{avg}}\times9/11)$, where $N_{\text{avg}}$ is the number
sub-segments whose periodograms are averaged together
in estimating the power spectrum for the optimal
filter.  For the data analyzed with the sliding power spectrum
estimator, 29 overlapping four-second sub-segments are averaged from
each of two 60-second segments, for a total $N_{\text{avg}}=58$.  This
gives a correction factor of
$1+1/(58\times9/11)=1.021$.  The ``na\"{\i}ve'' estimated sigmas, derived
from power spectra calculated using 29 averages
in a single 60-second segment, are scaled up by a
factor of $1+1/(29\times9/11)=1.042$.

\subsection{Combination of Analysis Segment Results}
\label{ss:post-combine}

As shown in \cite{Allen:1999}, the optimal way to combine a series of
independent cross-correlation measurements $\{Y_I\}$ with associated
one-sigma errorbars is 
\begin{subequations}
  \begin{gather}
    Y^{\text{opt}} = \frac{\sum_{I} \sigma_I^{-2} Y_I}
    {\sum_{I} \sigma_I^{-2}} \\
    \sigma_{Y^{\text{opt}}} = \left({\sum_{I} \sigma_I^{-2}}\right)^{-1/2}
  \end{gather}
\end{subequations}

To minimize spectral leakage, we use Hann windows in our analysis
segments, which would reduce the effective observing time by
approximately one-half, so we overlap the segments by 50\% to make
full use of as much data as possible.  This introduces correlations
between overlapping data segments, which modifies the optimal
combination slightly, as detailed in \cite{ovlpwin}.

\subsection{Statistical Interpretation}
\label{ss:post-bayes}

The end result of the analysis and post-processing of a set of data is
a an optimally combined complex cross-correlation statistic
$Y^{\text{opt}}$ with a theoretical mean of $\Omega_R T$ and an
associated standard deviation of $\sigma_{Y^{\text{opt}}}$ on both the
real and imaginary parts.  We can construct from this our overall
point estimate on the unknown actual value of $\Omega_R$ and the
corresponding one-sigma errorbar:
\begin{subequations}
  \begin{gather}
    \widehat{\Omega}_R = Y^{\text{opt}} / T \\
    \sigma_\Omega = \sigma_{Y^{\text{opt}}} / T
  \end{gather}
\end{subequations}
For a given value of $\Omega_R$, and assuming $\sigma_\Omega$ to be
given, the likelihood function for the complex point estimate to have
a particular value $\widehat{\Omega}_R = x + iy$ is
\begin{equation}
  \begin{split}
  P(x,y|\Omega_R,\sigma_{\Omega})
  &
  = \frac{d^2P}{dx\,dy}
  = \frac{1}{2\pi\sigma_{\Omega}^2}
  \exp\left(-\frac{\abs{x+iy-\Omega_R}^2}{2\sigma_{\Omega}^2}\right)
  \\
  &
  = \frac{1}{2\pi\sigma_{\Omega}^2}
  e^{-(x-\Omega_R)^2/2\sigma_{\Omega}^2}e^{-y^2/2\sigma_{\Omega}^2}    
  \end{split}
\end{equation}
Given a prior probability density function on $\Omega_R$, Bayes's
theorem allows us to construct a posterior
\begin{equation}
  \begin{split}
  P(\Omega_R|x,y,\sigma_{\Omega})
  &
  = \frac{P(x,y|\Omega_R,\sigma_{\Omega})P(\Omega_R)}
  {P(x,y|\sigma_{\Omega})}
  \\
  &
  \propto e^{-(x-\Omega_R)^2/2\sigma_{\Omega}^2}P(\Omega_R)
  \end{split}
\end{equation}
where $x=\Real\widehat{\Omega}_R$.  In this work we choose a uniform prior
over the interval $[0,\Omega_{\text{max}}]$, where
$\Omega_{\text{max}}$ is chosen to be 116 (the previous best upper
limit at frequencies around 900\,Hz \cite{Astone:1999}), except in the
case of the hardware injections in Sec.~\ref{ss:inj-hard}, where the a
priori upper limit is taken to be well above the level of the
injection.

Given a posterior probability density function (PDF), the 90\%
confidence level Bayesian upper limit $\Omega_{\text{UL}}$ is defined
by
\begin{equation}
  \int_0^{\Omega_{\text{UL}}} d\Omega_R\,P(\Omega_R|x,y,\sigma_{\Omega}) = 0.90
\end{equation}
Alternatively, any range containing 90\% of the area under the
posterior PDF can be thought of as a Bayesian 90\% confidence level
range.  To allow consistent handling of results with and without
simulated signals, we choose the narrowest range which represents 90\%
of the area under the posterior PDF.  (This is the range whose PDF
values are larger than all those outside the range.)  For a low enough
signal-to-noise ratio $\Real\widehat{\Omega}_R / \sigma_{\Omega}$,
this range is from 0 to $\Omega_{\text{UL}}$.

\subsection{Treatment of Calibration Uncertainty}
\label{ss:post-cal}

In reality, the conversion of raw data from the LIGO and ALLEGRO GW
detectors into GW strain is not perfect.  We model this uncertainty in
the calibration process as a time- and frequency-independent phase and
magnitude correction, so that $\widehat{\Omega}_R=x+iy$ and
$\sigma_{\Omega}$ are actually related to $\Omega_R
e^{\Lambda+i\phi}$, where $\Lambda$ and $\phi$ are unknown amplitude
and phase corrections; the likelihood function then becomes
\begin{equation}
  P(x,y|\Omega_R,\sigma_{\Omega},\Lambda,\phi)
  = \frac{1}{2\pi\sigma_{\Omega}^2}
  \exp\left(-\frac{\abs{x+iy-\Omega_Re^{\Lambda+i\phi}}^2}
    {2\sigma_{\Omega}^2}\right)
\end{equation}
Given a prior PDF $P(\Lambda,\phi)$ for the calibration corrections,
we can marginalize over these nuisance parameters and obtain a 
marginalized likelihood function
\begin{multline}
  P(x,y|\Omega_R,\sigma_{\Omega})
  \\
  =\int_{-\infty}^{\infty} d\Lambda \int_{-\pi}^{\pi} d\phi
  \, P(x,y|\Omega_R,\sigma_{\Omega},\Lambda,\phi) \, P(\Lambda,\phi)
\end{multline}
We take this prior PDF to be Gaussian in $\Lambda$ and $\phi$, with a
standard deviation added in quadrature from the quoted amplitude and
phase uncertainty for the two instruments.  For L1, this is 5\% in
amplitude and $2^\circ$ in phase \cite{s4cal} and for A1
it is 10\% in amplitude and $3^\circ$ in phase \cite{s4ALLEGROcal}.

\section{ANALYSIS OF COINCIDENT DATA}
\label{s:res}

\subsection{Determination of Analysis Parameters}
\label{ss:res-params}

\begin{figure}[htbp]
  \centering
  \includegraphics[width=240pt]{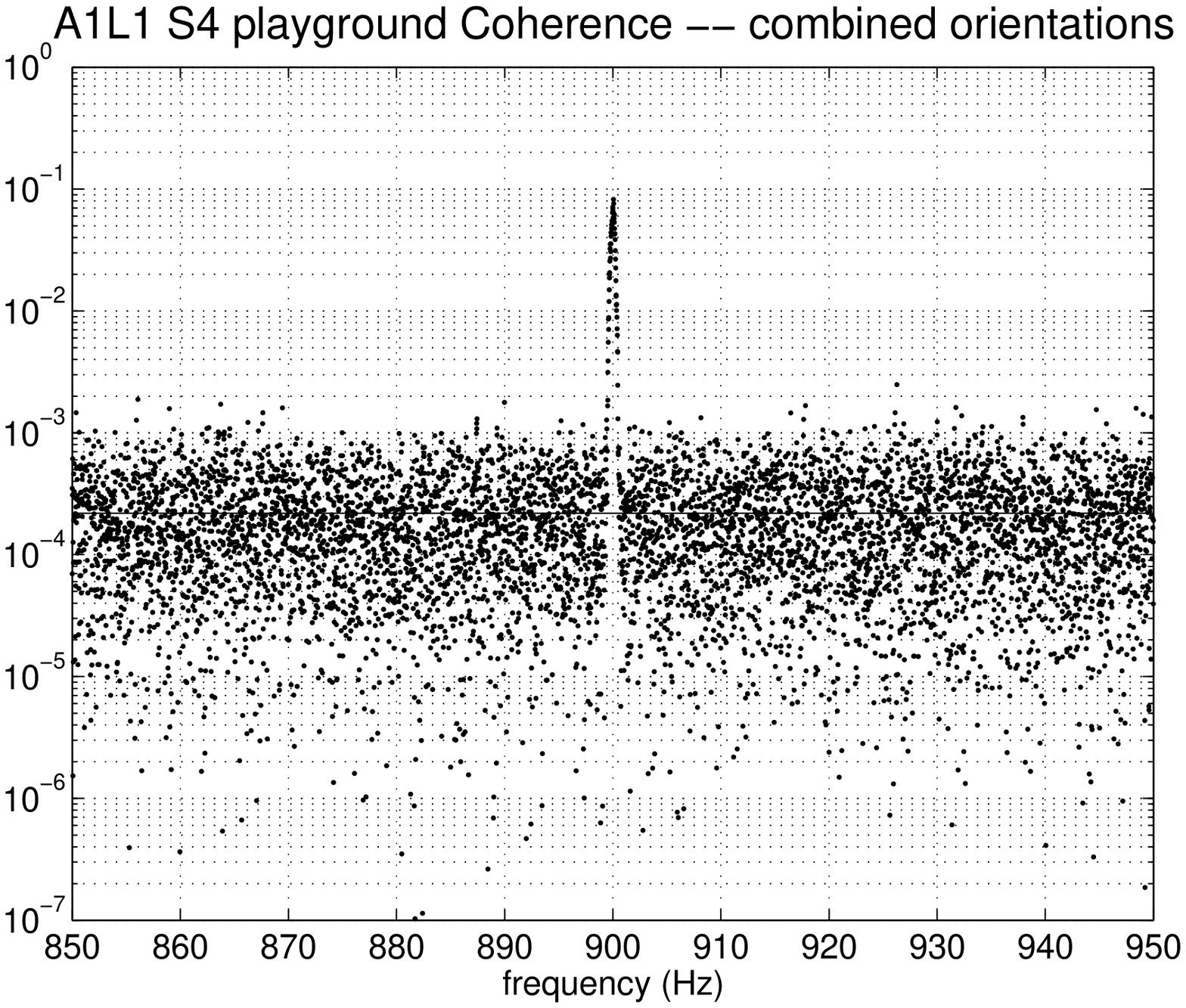}
  \includegraphics[width=240pt]{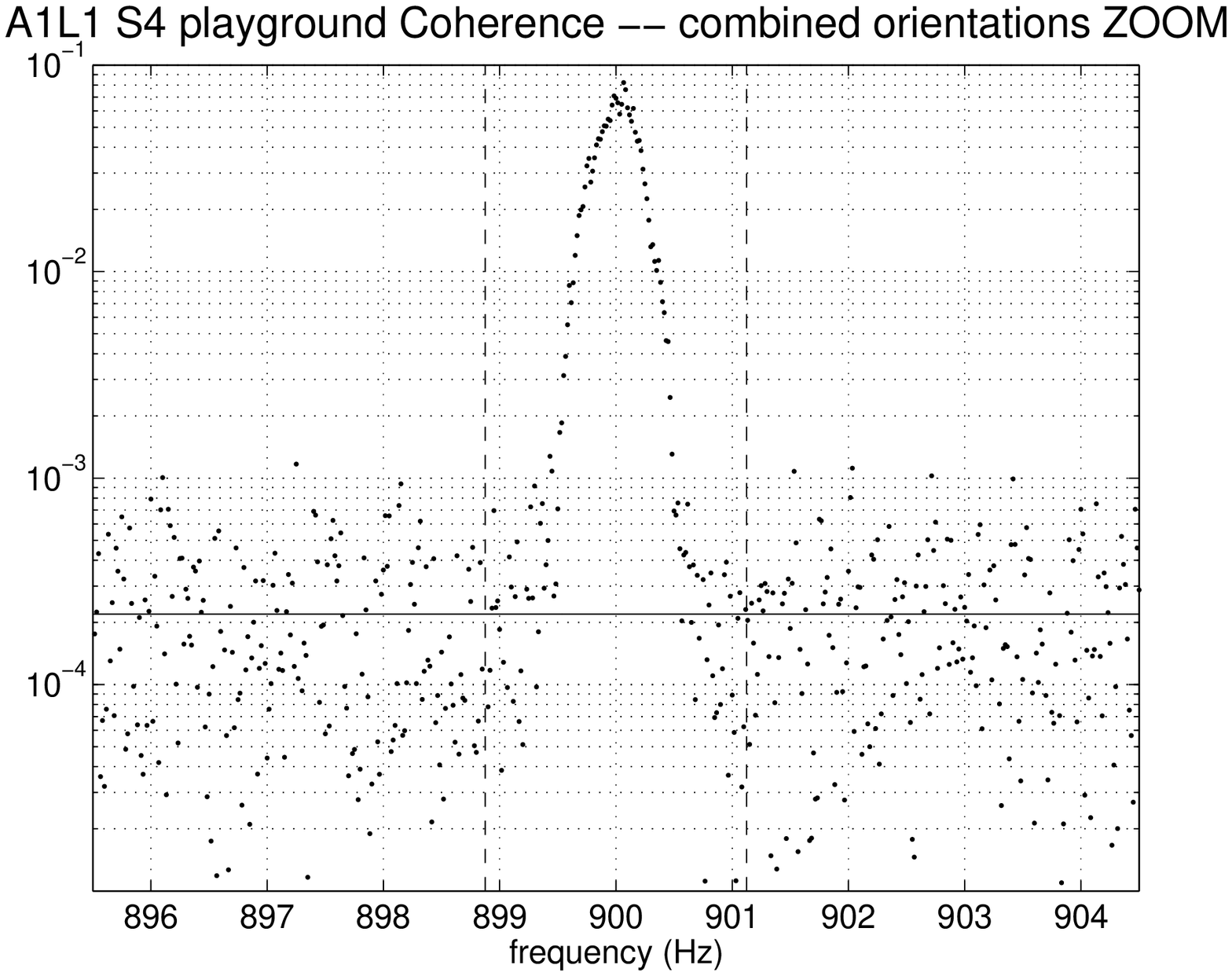}
  \caption{LLO-ALLEGRO (L1-A1) coherence, calculated from 48.66~hours
    of playground data spanning nearly 30 days.  The only significant feature is the power
    line harmonic at 900\,Hz.  The closeup view in the second plot
    shows that the coherence is insignificant beyond 1\,Hz away from
    the line.  Based on this, we mask out the nine 0.25-Hz frequency
    bins around 900\,Hz from our analysis.}
\label{fig:coherence}
\end{figure}

\begin{figure}[htbp]
  \centering
  \includegraphics[width=240pt]{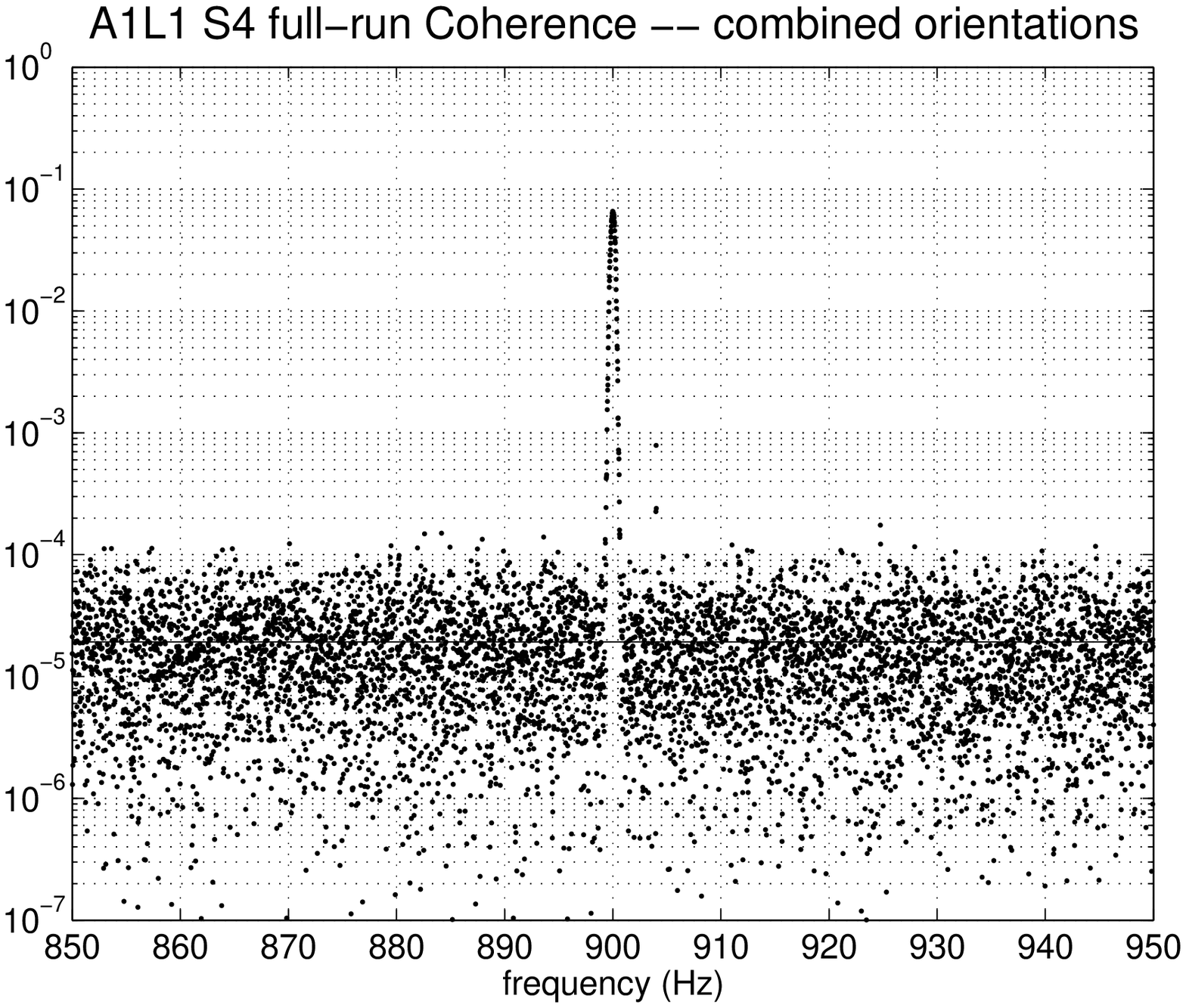}
  \includegraphics[width=240pt]{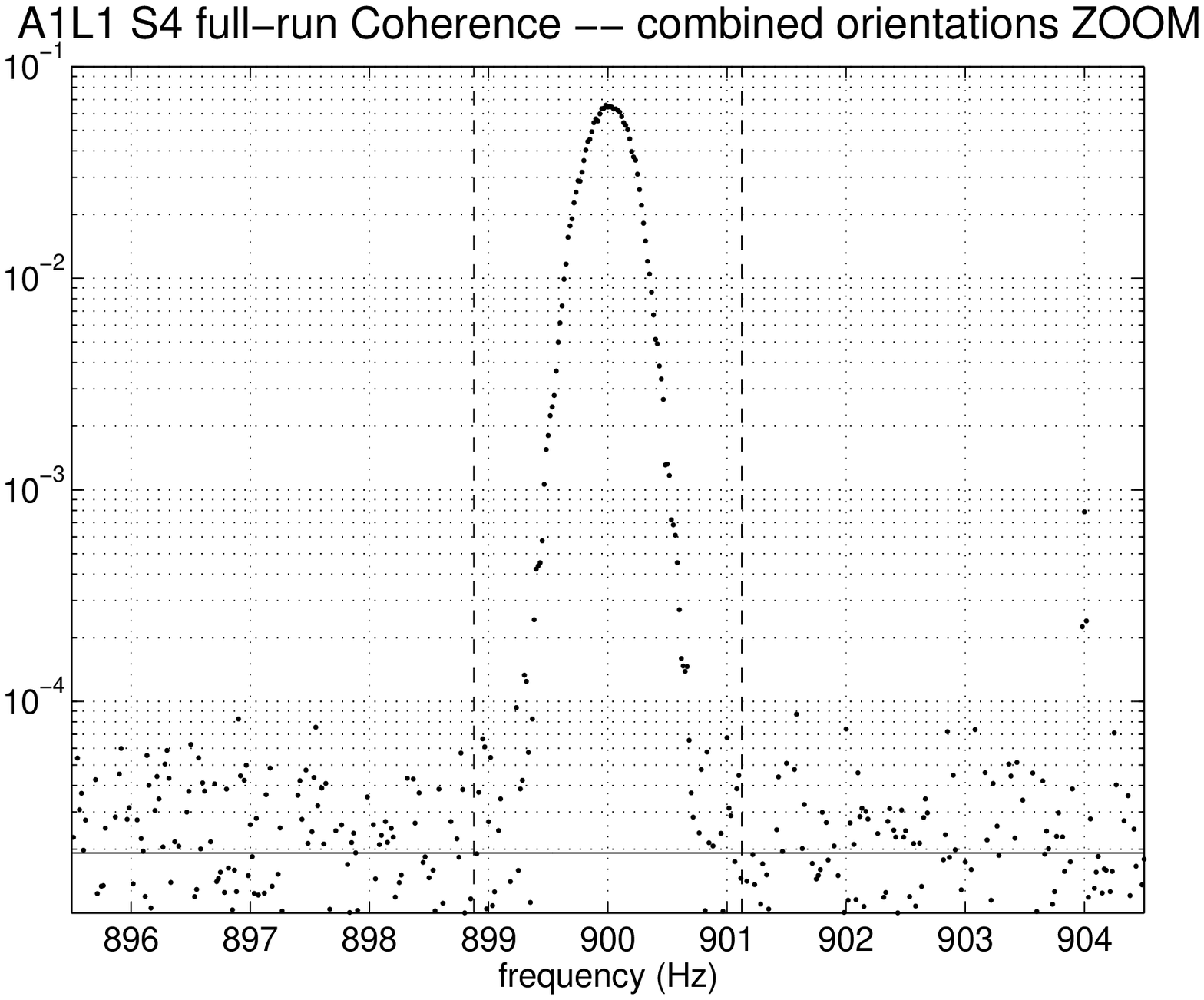}
  \caption{LLO-ALLEGRO (L1-A1) coherence, calculated from all S4 data without
    regard to playground status.  Again, the 900\,Hz line is seen to
    be comfined to a 2\,Hz wide range.  Additionally, a feature at the
    heterodyning frequency of 904\,Hz (which was masked \textit{a
      priori} in our main cross-correlation analysis) becomes
    visible.}
\label{fig:coherence_ALL}
\end{figure}

To avoid biasing our results, we set aside approximately 9\% of the
data, spread throughout the run, as a \textit{playground} on which to
tune our analysis parameters.  Based on playground investigations, we
settled on the following parameters for our analysis:
\begin{itemize}
\item Overlapping 60-second Hann-windowed analysis segments
\item Frequency range $850\un{Hz}\le f\le 950\un{Hz}$, $0.25\un{Hz}$
  resolution for optimal filter
\item L1 data downsampled to 4096\,Hz before analysis
\item Removal of the following frequencies from the optimal filter:
  $900\un{Hz}$ ($2.25\un{Hz}$ wide), $904\un{Hz}$ ($0.25\un{Hz}$
  wide).
\end{itemize}
The frequencies to remove were chosen on the basis of studies of the
coherence of stretches of L1 and A1 data
(see Fig.~\ref{fig:coherence}); $900\un{Hz}$ is the
only harmonic of the 60\,Hz power line within our analysis band, and
detectable coherence is seen for frequencies within $1\un{Hz}$ of the
line.  $904\un{Hz}$ is notched out because, as the heterodyne
frequency, it corresponds to DC in the heterodyned A1 data.
After completing the cross-correlation analysis, we computed the
coherence from the full run of data, as shown in
Fig.~\ref{fig:coherence_ALL}.  The results are similar to those from
the playground, except for a lower background level, and a feature at
the heterodyning frequency.

\begin{figure}[htbp]
  \centering
  \includegraphics[width=240pt]{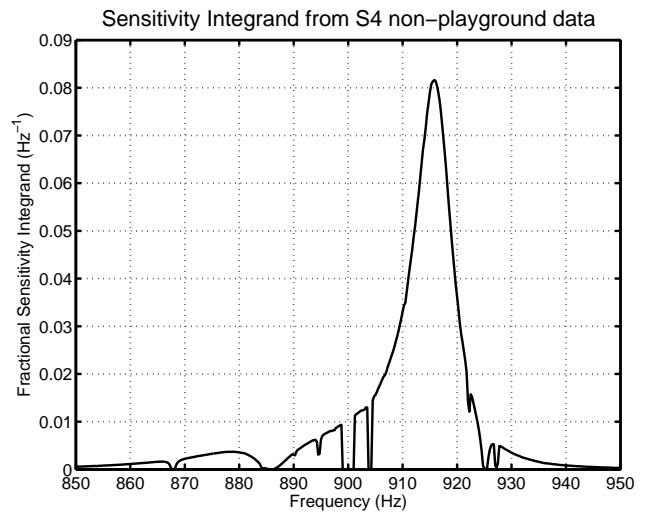}
  \caption{The sensitivity integrand for the data used in the
    cross-correlation analysis, normalized so its integral equals
    unity.  The area under this curve, between two frequencies, is
    the fractional contribution to $\sigma^{-2}$ from that range of
    frequencies.  Notice that the nine frequency bins masked out
    around 900\,Hz and the one at 904\,Hz give no contribution, and
    that the sensitivity integrand is also suppressed at other
    frequencies corresponding to lines in the A1 noise power
    spectrum.}
\label{fig:tot_sens}
\end{figure}

The relevant range of frequencies can be determined by looking at the
support of the integrand of \eqref{e:sigmatheor}, known as the
\textit{sensitivity integrand}.  The overall sensitivity integrand,
constructed as a weighted average over all the non-playground data
used in the analysis, is plotted in Fig.~\ref{fig:tot_sens}.
The area under this curve for a range of frequencies is proportional
to that frequency range's contribution to $\sigma^{-2}$.  We see that
the integrand does indeed become negligible by a frequency of 850\,Hz
on the lower end and 950\,Hz on the upper end.  We further see that
most of the sensitivity comes from a 20-Hz wide band centered around
915\,Hz.

\subsection{Cross-Correlation and Upper Limit Results}
\label{ss:res-res}

After data quality cuts, exclusion of the ``playground'', and
application of the stationarity cut described in
Sec.~\ref{ss:post-sigmarat}, {\numSegs} one-minute segments of data
were analyzed, for an effective observing time of {\effHrs} hours
(considering the effects of Hann windowing), of which
{\effHrsXARM} hours was in
the XARM orientation, {\effHrsYARM} in the YARM orientation,
and {\effHrsNULL}
in the NULL orientation.  The results are shown in
Table~\ref{tab:S4Res}.  No statistically significant correlation is
seen in any orientation, and optimal combination of all data leads to
a point estimate of $\ptEst$ for $\Omega_R$, with a
one-sigma errorbar of $\eBar$ each on the real and imaginary
parts.
\begin{table}
    \begin{tabular}{||l||c||c|c||}
    \hline
    \hline
    & $T_{\text{eff}}$ & \multicolumn{2}{c||}{$\Omega_R$}\\
Type&(hrs)&Point Estimate&Error Bar\\
    \hline
    \hline
\qquad{XARM}&{181.2}&{$0.61 + 0.25i$}&{0.56}\\
    \hline
\qquad{YARM}&{114.7}&{$-0.47 + 0.47i$}&{0.90}\\
    \hline
    \hline
\quad{non-NULL}&{295.8}&{$0.31 + 0.31i$}&{0.48}\\
    \hline
    \hline
\qquad{NULL}&{88.2}&{$10.96 - 43.89i$}&{28.62}\\
    \hline
    \hline
all&{384.1}&$0.31 + 0.30i$&0.48\\
    \hline
    \hline
  \end{tabular}

  \caption{
    Results of optimally-filtered cross-correlation of non-playground
    data.  Results are shown for data in each of three orientations (XARM,
    YARM, and NULL).  Additionally, the XARM and YARM results are combined
    with the optimal weighting (proportional to one over the square of the
    errorbar) to give a ``non-NULL'' result, and results from all three
    orientations are optimally combined to give an overall result.
    In each case, $T_{\text{eff}}$ is the effective observing time
    including the effects of overlapping Hann windows.
    Note
    that since the non-NULL data are much more sensitive than the NULL
    data, they dominate the final result.  Note also that because the
    optimal filter includes the the ORF, the
    relative orientation of LLO and ALLEGRO is already included in these
    results.  This is reflected, for example, in the large errorbars on
    the NULL result.
  }
  \label{tab:S4Res}
\end{table}

The results in Table~\ref{tab:S4Res} include the ORF describing the
geometry in the optimal filter.  This means an orientation-independent
non-GW cross-correlation present in the data would change sign between
XARM and YARM, and would look much larger in the NULL result.  One way
to remove the effects of the observing geometry and compare
orientation-independent cross-correlations is to remove the
$\gamma(f)$ from \eqref{eq:Qoptnorm}.  Since the ORF for each
orientation is nearly constant across the observing band, and notably
across the region of peak sensitivity, it is sufficient to multiply
the overall results in each case by $\gamma(915\un{Hz})$.  This is
shown in Table~\ref{tab:S4GammaRes}, where we again see no significant
cross-correlation, and sensitivities whose relative sizes are well
explained by the differing observing times.
\begin{table}
    \begin{tabular}{||l||c||c|c||}
    \hline
    \hline
    & $T_{\text{eff}}$ & \multicolumn{2}{c||}{$\gamma\Omega_R$}\\
Type&(hrs)&Point Estimate&Error Bar\\
    \hline
    \hline
{XARM}&{181.2}&{$0.58 + 0.24i$}&{0.53}\\
    \hline
{YARM}&{114.7}&{$0.42 - 0.42i$}&{0.80}\\
    \hline
{NULL}&{88.2}&{$0.35 - 1.40i$}&{0.91}\\
    \hline
    \hline
  \end{tabular}

  \caption{
    The cross-correlation results of Table~\protect\ref{tab:S4Res},
    scaled by $\gamma(915\un{Hz})$ from
    Table~\protect\ref{tab:orientations}, the value
    of the ORF at the frequency of peak sensitivity.  This gives a sense
    of the ``raw'' cross-correlation, independent of the
    orientation-dependent geometrical factor.  The different observing
    times explain the remaining variation in the one-sigma errorbar for
    the measurement, which should be inversely proportional to the square
    root of the observing time.
  }
  \label{tab:S4GammaRes}
\end{table}

\begin{figure}[htbp]
  \centering
  \includegraphics[width=240pt]{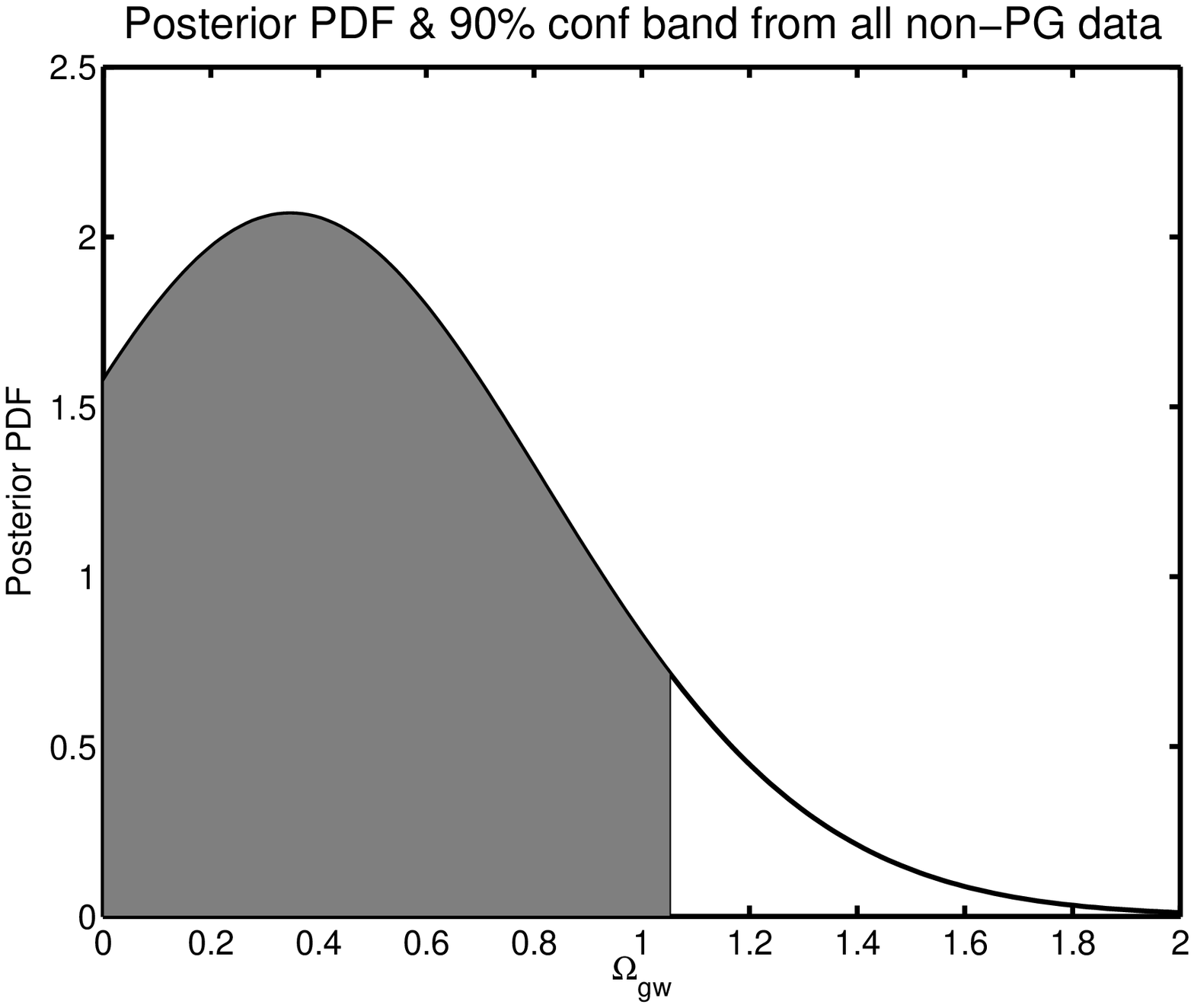}
  \caption{Posterior probability density function associated with the
    overall combined point estimate of $\ptEst$ and
    estimated errorbar of $\eBar$, considering the uncertainty
    in the phase and amplitude of the calibration.  The shaded region
    represents 90\% of the area under the curve, leading to an upper
    limit on $\Omega_R$ of $\omegaLim$, which corresponds to
    a gravitational wave strain of $\strainLim$ at the peak frequency
    of $915\un{Hz}$.}
\label{fig:NPG_postrange}
\end{figure}

We use the methods of Sec.~\ref{ss:post-bayes} and
\ref{ss:post-cal} to construct a posterior PDF from the overall
cross-correlation measurement of $\ptEst$ and estimated
errorbar of $\eBar$, taking into consideration the nominal
calibration uncertainty of {11\%} in magnitude and
{$3.6^\circ$} in phase to obtain the posterior PDF shown
in Fig.~\ref{fig:NPG_postrange}.
The narrowest likely 90\% confidence interval on $\Omega_R$ is
$[0,\omegaLim]$.  We thus set an upper limit of $\omegaLim$ on
$\Omega_R=\Omega_{\text{gw}}(f_R)$, which translates to an upper limit on
$S_{\text{gw}}(915\un{Hz})$ of $(\strainLim)^2$.

\section{VALIDATION VIA SIGNAL INJECTION}
\label{s:inj}

To check the effectiveness of our algorithm at detecting stochastic GW
signals, we performed our search on data with simulated waveforms
``injected'' into them.  This was done both by introducing the
simulated signals into the analysis pipeline (software injections),
and by physically driving both instruments in coincidence (hardware
injections).
Hardware injections provide
an end-to-end test of our detection pipeline and also a test
on the calibration accuracies of our instruments,
but are necessarily short in duration because they corrupt the GW data
taken during the injection.  Software injections can be carried out
for longer times and therefore at lower signal strengths, and can be
repeated to perform statistical studies.
Software injections,
however, cannot check for calibration errors.

\subsection{Signal Simulation Algorithm}
\label{ss:inj-alg}

To simulate a correlated SGWB signal in an interferometer and a bar,
the formulas in, e.g., \cite{Allen:1999} need to be generalized
slightly.  This is because the ORF of a
detector with itself is in general \cite{Whelan:2006}
\begin{equation}
  \gamma = 2\left[d^{ab}d_{ab} - \frac{1}{3} (d^a_a)^2\right] \ ,
\end{equation}
which is unity for an IFO with perpendicular arms \eqref{e:difo} 
but 4/3 for a bar \eqref{e:dbar}. Writing this quantity for 
detector 1 or 2 as $\gamma_{11}$ or $\gamma_{22}$, respectively,
makes the required
cross-correlations in a simulated SGWB signal
\begin{subequations}\label{strainCovar}
  \begin{align}
    \langle \widetilde{h}_1^*(f) \widetilde{h}_1(f') \rangle
    &= \frac{1}{2} \delta(f-f')\,S_{\text{gw}}(f)\,\gamma_{11}
    \\
    \langle \widetilde{h}_1^*(f) \widetilde{h}_2(f') \rangle
    &= \frac{1}{2} \delta(f-f')\,S_{\text{gw}}(f)\,\gamma_{12}(f)
    \\
    \langle \widetilde{h}_2^*(f) \widetilde{h}_2(f') \rangle
    &= \frac{1}{2} \delta(f-f')\,S_{\text{gw}}(f)\,\gamma_{22} \,.
  \end{align}
\end{subequations}
The above expressions do not determine a unique algorithm for
converting a set of random data streams into individual detector
strains. One possible prescription is
\begin{subequations}\label{e:h1h2sim}
  \begin{align}
    \widetilde{h}_1(f)
    =&\ 
    \frac{1}{2} \sqrt{S_{\text{gw}}(f)}~\sqrt{\gamma_{11}}
    \left(x_{1}(f) + i y_{1}(f)\right)
    \\
    \begin{split}
      \widetilde{h}_2(f)
      =&\ 
      \widetilde{h}_1(f)\frac{\gamma_{12}(f)}{\gamma_{11}} 
      \\
      &+
      \frac{1}{2} \sqrt{S_{\text{gw}}(f)
        \left(\gamma_{22}-\frac{\gamma_{12}^2(f)}{\gamma_{11}}\right)}\
      (x_{2}(f) + i y_{2}(f))
      \ ,
    \end{split}
  \end{align}
\end{subequations}
where $x_{1}(f)$, $y_{1}(f)$, $x_{2}(f)$, and $y_{2}(f)$ are statistically 
independent real Gaussian random variables, each of zero mean and unit 
variance.  In the above pair, $\gamma_{12}(f)$ is used only in the
construction of $\widetilde{h}_2(f)$ and not of $\widetilde{h}_1(f)$.
A different pair, where $\gamma_{12}$ is explicitly included in the
calculation of both strains more symmetrically, can be defined as
follows: Let $z_k(f):= (x_k(f) + iy_k(f))/\sqrt{2}$ be a pair
($k=1,2$) of complex random functions and let $s :=
\sqrt{1-\gamma_{12}^2/(\gamma_{11}\gamma_{22})}$. Then, the second
pair can be expressed as:
\begin{subequations}\label{e:h1h2simSym}
  \begin{align}
    \widetilde{h}_1(f)
    &=
    \sqrt{\frac{S_{\text{gw}}(f)}{2}}~\sqrt{\gamma_{11}}
    \left(a(f)z_{1}(f) + b(f)z_{2}(f)\right)
    \\
    \widetilde{h}_2(f)
    &=
    \sqrt{\frac{S_{\text{gw}}(f)}{2}}~\sqrt{\gamma_{22}}
    \left(b(f)z_{1}(f) + a(f)z_{2}(f)\right)
    \ , 
  \end{align}
\end{subequations}
where $a=\sqrt{(1+s)/2}$ and
$b=\gamma_{12}/\sqrt{2(1+s)\gamma_{11}\gamma_{22}}$ are determined
completely by the three ORFs.
Either pair of simulated strains obeys \eqref{strainCovar}.
The signals for software injections were generated using
\eqref{e:h1h2simSym}; those for hardware injections were generated
by an older code which used \eqref{e:h1h2sim}.
Further details of simulated signal generation
are in \cite{Bose:2003nb}.

\subsection{Results of Software Simulation}
\label{ss:inj-soft}

\begin{table}
    \begin{tabular}{||c||c||c||c||}
    \hline
    \hline
    $\Omega_R$ injected & Point Estimate & Error Bar & 90\% conf int\\    \hline
    \hline
    0 & $0.32 - 1.00i$ & 1.54 & [0.00,2.74]\\
    \hline
    1.9 & $2.22 - 0.86i$ & 1.55 & [0.09,4.35]\\
    \hline
    3.9 & $4.14 - 0.79i$ & 1.55 & [1.61,6.66]\\
    \hline
    9.6 & $9.89 - 0.65i$ & 1.56 & [7.32,12.45]\\
    \hline
    19 & $19.56 - 0.49i$ & 1.58 & [16.96,22.15]\\
    \hline
    \hline
  \end{tabular}

  \caption{Results of software injections.  All figures are for
    constant-$\Omega_{\text{gw}}(f)$ and listed by
    $\Omega_R$ level.  The 90\% confidence
    level ranges are calculated without marginalizing over any
    calibration uncertainty.}
  \label{tab:inj-soft}
\end{table}  
We performed software injections into the full S4 coincident
playground, {\numSegsSW} overlapping one-minute analysis segments with
an effective observing time of {\effHrsSW} hours considering the
effects of Hann windows ({\effHrsSWXARM} hours of this is in the XARM
orientation, {\effHrsSWYARM} hours in the YARM orientation,
and {\effHrsSWNULL} hours in the
NULL orientation).  We injected constant-$\Omega_{\text{gw}}(f)$
spectra of strengths corresponding to
$\Omega_R=${1.9, 3.9, 9.6, and 19}, as well as a test
with the SGWB amplitude set to 0 to reproduce the analysis of the
playground itself.  Note that even the strongest of these injections
does not produce correlations detectable in an individual one-minute
analysis segment.  The results are summarized in
table~\ref{tab:inj-soft}.  In each case, the actual injected value of
$\Omega_R$ is consistent with the real part of
the point estimate to within the one-sigma estimated error bar; the
imaginary part of the point estimate 
remains zero to within the errorbar.
The results for injections at
{$\Omega_R=1.9$} and stronger would correspond to
statistical ``detections'' at the 90\% or better confidence level.

\begin{table*}
    \begin{tabular}{||c||c||c||c||c|c|c||c|c||c|c||}
    \hline
    \hline
    & $\gamma\Omega_R$ 
    & $\gamma\Omega_R$ 
    & $\Omega_R$ 
    & \multicolumn{3}{c||}{$\Omega_R$ Point Estimate}    & \multicolumn{2}{c||}{unmarg.~range}    & \multicolumn{2}{c||}{marg.~range}\\
Injection&Error Bar&Point Estimate&Error Bar&Value&Mag&Phase&min&max&min&max\\
    \hline
{A-minus}&{83}&$-6623 - 126i$&{93}&$7403 + 141i$&{7404}&$1.1^\circ$&{7250}&{7555}&{6212}&{8820}\\
    \hline
{A-null}&{99}&$205 + 19i$&{3106}&$6429 + 607i$&{6457}&$5.4^\circ$&{1565}&{11294}&{1435}&{11562}\\
    \hline
{A-plus}&{83}&$6983 + 64i$&{87}&$7325 + 67i$&{7325}&$0.5^\circ$&{7182}&{7468}&{6146}&{8726}\\
    \hline
{B-minus}&{95}&$-6845 + 49i$&{106}&$7650 - 55i$&{7651}&$-0.4^\circ$&{7478}&{7825}&{6417}&{9115}\\
    \hline
{B-null}&{111}&$366 - 50i$&{3486}&$11492 - 1576i$&{11600}&$-7.8^\circ$&{5950}&{17035}&{5857}&{17365}\\
    \hline
{B-plus}&{92}&$7128 + 77i$&{96}&$7477 + 80i$&{7477}&$0.6^\circ$&{7317}&{7634}&{6272}&{8907}\\
    \hline
    \hline
{all}& \multicolumn{2}{c||}{N/A}&{47}&$7448 + 65i$&{7448}&$0.5^\circ$&7371&7526&6256&8867\\
    \hline
    \hline
  \end{tabular}

  \caption{
    Results of hardware injections.  Simulated waveforms with an effective
    signal strength of $\Omega_{\text{gw}}(f)=\HWinj$ were injected
    coincidentally in the ALLEGRO and the LIGO-Livingston (LLO) detectors.
    The ``A'' and ``B'' sets of injections took place during the XARM and
    NULL observing times, respectively.  Independent of the actual
    orientation, the simulated signals were generated and analyzed
    assuming different orientations: YARM, NULL, and XARM orientations
    were assumed for the injections labelled ``minus'', ``null'', and
    ``plus'', respectively.
    The first pair of columns shows the errorbars and point estimates
    scaled by $\gamma(915\un{Hz})$ to give a ``raw'' cross-correlation
    as in Table~\protect\ref{tab:S4GammaRes}.
    (Note that since the ``null'' alignment represents
    not-quite-perfect misalignment, as noted in
    Table~\protect\ref{tab:orientations} [$\gamma(915\un{Hz})=0.03$ rather
    than zero], the injection still leads to a statistically significant
    cross-correlation even in the ``null'' orientation.)  Note that the
    errorbars, thus scaled, are comparable for all six injections, while
    the level of correlation or anti-correlation depends on the
    orientation associated with the injection being analyzed.
    The subsequent columns relate to the standard cross-correlation
    statistic, with the ORF included in the optimal filter, so the
    relative insensitivity in the ``null'' alignment is reflected
    in large errorbars, while the point estimates are all positive
    and in the vicinity of the injected value of $\HWinj$.
    All point estimates given include corrections for known
    phase offsets associated with the injection system.  The row labelled
    ``all'' gives the optimally-weighted combination of all six results.
    The point estimates and one-sigma estimated errorbars were combined
    to give 90\% confidence ranges with and without marginalization over
    calibration uncertainty, mimicking the statistical analysis described
    in Sec.~\protect\ref{ss:post-bayes} and~\protect\ref{ss:post-cal}.
  }
  \label{tab:inj-hard}
\end{table*}

\begin{figure*}[htbp]
  \centering
  \includegraphics[width=240pt]{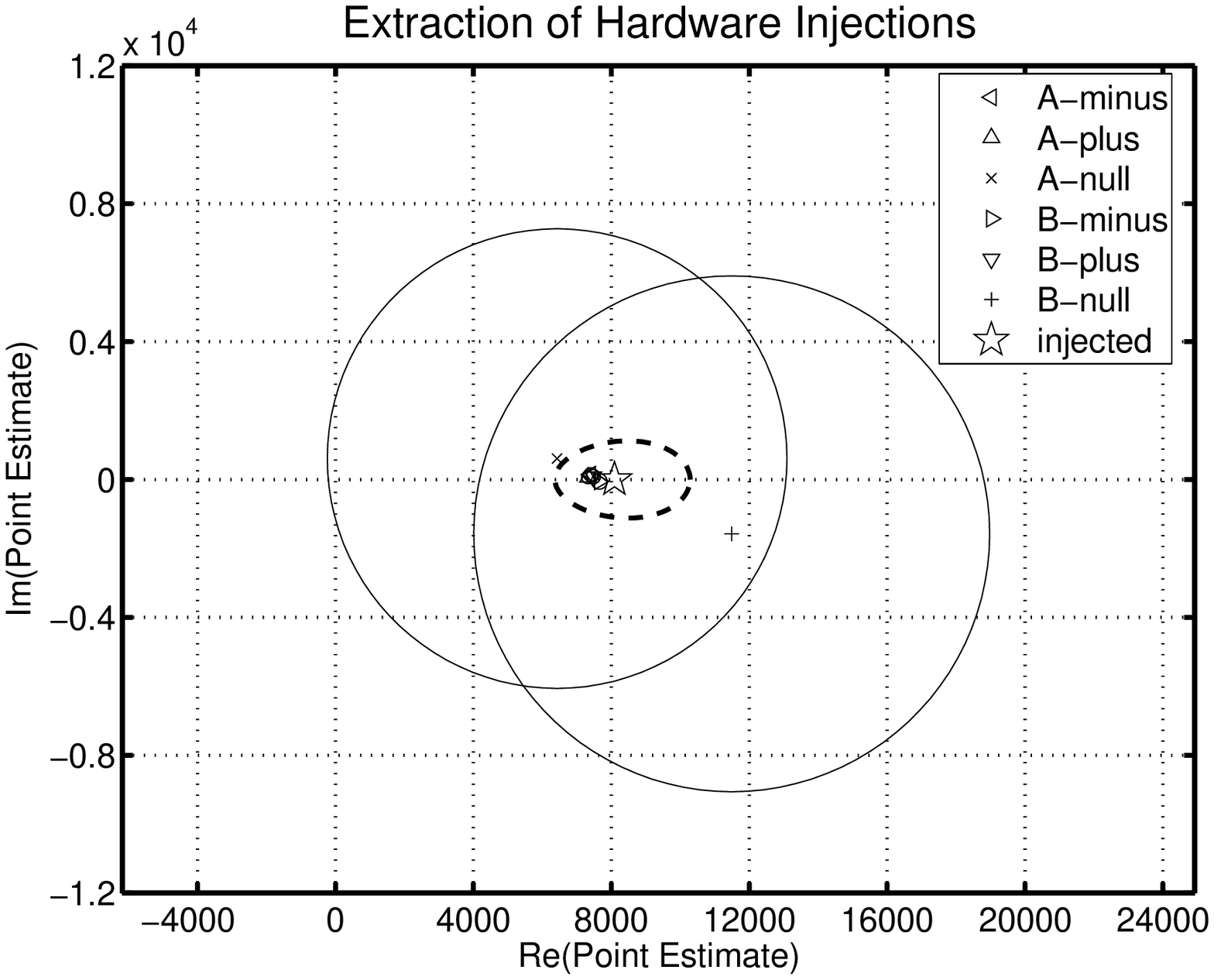}
  \includegraphics[width=240pt]{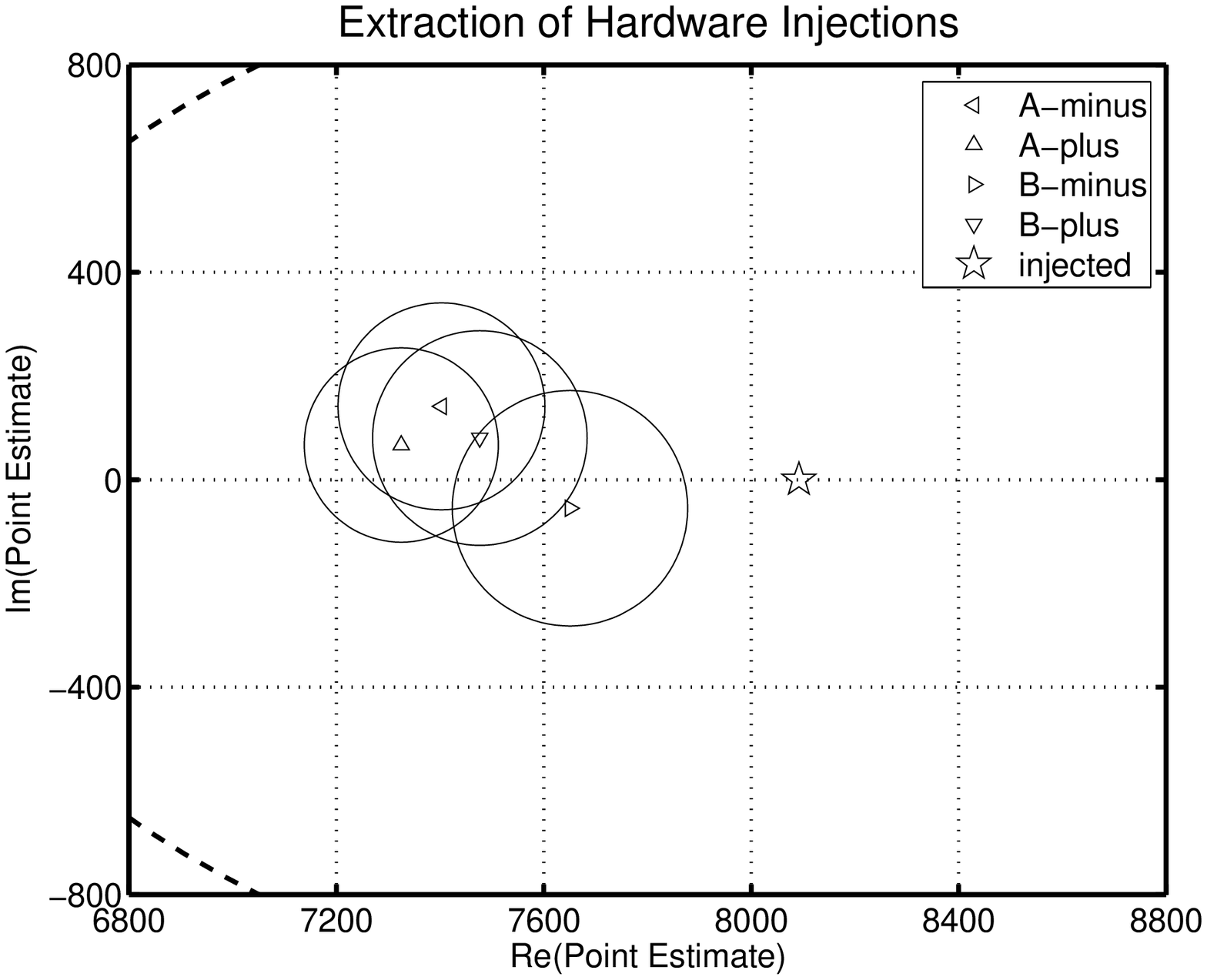}
  \caption{Visualization of hardware injection results.  Each of the
    six point estimates of $\Omega_R$ is plotted
    on the complex plane, with an associated error circle of 2.15
    times the estimated one-sigma errorbar.  (This contains 90\% of
    the volume under the corresponding likelihood surface.)  The
    five-pointed star indicates the actual injected level of
    $\Omega_R=\HWinj$.  The dashed teardrop-shaped
    region indicates the calibration uncertainty, corresponding to a
    2.15-sigma ellipse in log-amplitude/phase space.  On the left we
    see that the two ``null'' injections are consistent in amplitude
    and phase with the actual injection, considering the statistical
    uncertainty associated with the real and imaginary parts of their
    point estimates.  The plot on the right (in which the edge of the
    dashed calibration uncertainty teardrop can just be seen) shows that
    the ``plus'' and ``minus'' injections are all statistically
    consistent with each other, and consistent with the injection when
    systematic uncertainties associated with calibration are taken
    into account.}
\label{fig:HWplane}
\end{figure*}

\begin{figure}[htbp]
  \centering
  \includegraphics[width=240pt]{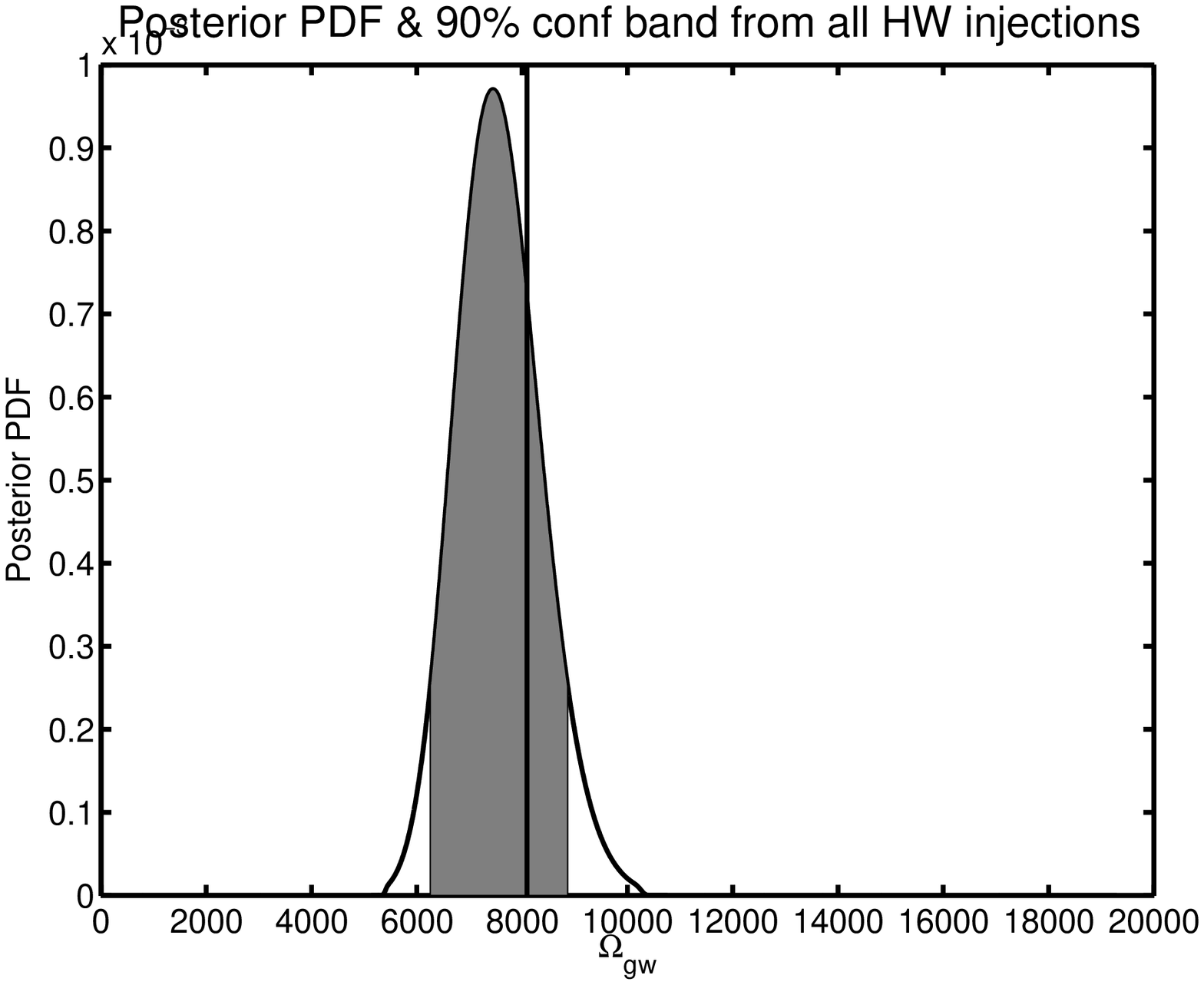}
  \caption{Posterior probability density function associated with
    combined hardware injection measurement, including marginalization
    over calibration uncertainties.  Note that while the combined
    one-sigma statistical errorbar is only {\eBarHW}, the shaded 90\%
    area under the curve has a width of over {2000}.  This is because
    systematic errors dominate in the presence of the large point
    estimate.  The solid vertical line indicates the actual injection
    level of $\Omega_R=\HWinj$.}
\label{fig:postrange_HW.eps}
\end{figure}

\subsection{Results of Hardware Injection}
\label{ss:inj-hard}

During S4, a set of simulated signals was injected in the hardware of 
ALLEGRO and LLO. These injections served to test the full detection pipeline
as well as the calibrations of both instruments.
As described in detail in \cite{Bose:2003nb}, the preparation 
of simulated waveforms for hardware injections requires application of
the transfer function of the hardware component that is actuated, 
such as one of the two end test masses in LLO, to the theoretical strain
for that instrument.  Subsequent refinements to instrumental
calibration mean that the precise injected signal strength is
determined after the fact.  Six hardware injections performed during
S4, each 1020 seconds long, had an effective constant
$\Omega_{\text{gw}}(f)$ of {\HWinj}.  The series of injections we
call ``A'' and ``B'' were performed during the XARM and NULL
observation periods, respectively.  Independent of the physical
orientation of ALLEGRO, the injection and analysis was performed for
three different assumed orientations, producing ``plus'' (aligned, as
in the XARM orientation), ``minus'' (anti-aligned, as in YARM), and
``null'' (misaligned, as in NULL) injections in each series.  The
results (after correcting for known phase offsets in the injection
systems) for each of the six injections are shown in
Table~\ref{tab:inj-hard}.

The results show some variation of magnitude and phase of the point
estimates, especially for the ``null'' injections.  However, all the
injection results are consistent with the injected signal strength to
within statistical and systematic uncertainties.  This is illustrated
informally in Fig.~\ref{fig:HWplane}, which shows the point estimates
on the complex plane, each surrounded by an error circle of radius
2.15 times the corresponding estimated errorbar.  (This radius was
chosen because 90\% of the volume under a two-dimensional Gaussian
falls within a circle of radius $2.15\sigma$.)  Those circles all
overlap with a region centered at the actual injection strength
illustrating the magnitude and phase uncertainty in the calibration.
The systematic error can be more quantitatively evaluated using the
method of Sec.~\ref{ss:post-cal} to produce a posterior PDF associated
with each injection measurement.  This is illustrated for the
optimally-combined point estimate of {$\ptEstHW$} and associated
estimated errorbar {\eBarHW} in Fig.~\ref{fig:postrange_HW.eps}, and
ranges corresponding to the most likely 90\% confidence range under
the posterior PDFs (with and without marginalization) are included in
Table~\ref{tab:inj-hard}.  For each of the six injections, as well as
for the combined result, the actual injected value of {\HWinj} falls
into the range after marginalization over the calibration uncertainty.

\section{COMPARISON TO OTHER EXPERIMENTS}
\label{s:others}

The previous most sensitive direct upper limit at the frequencies
probed by this experiment was set by cross-correlating the outputs of
the EXPLORER and NAUTILUS resonant bar detectors \cite{Astone:1999}.
They found an upper limit on
$h_{100}^2\Omega_{\text{gw}}(907.20\un{Hz})$ of 60.  Using the value
of $72\un{km/s/Mpc}$ for the Hubble constant, that translates to a
limit of 116 on $\Omega_{\text{gw}}(907.20\un{Hz})$, upon which our
limit of $\omegaLim$ is a hundredfold improvement.

Data from LLO, taken during S4, were also correlated with data from
the LIGO Hanford Observatory (LHO) to set an upper limit on
$\Omega_{\text{gw}}(f)$ at frequencies between 50\,Hz and
150\,Hz \cite{s4sgwb}.  Correlations between LLO and LHO are not suited
to measurements at high frequencies because of the effects of the ORF,
illustrated in Fig.~\ref{fig:overlap}.  For comparison, rough measurements
using S4 LLO-LHO data and a band from 850\,Hz to 950\,Hz yield
upper limits of around {20}, while those confined to $905\un{Hz}\le
f\le925\un{Hz}$ (the band contributing most of the L1-A1
sensitivity give upper limits of around {80}.

Correlations between the 4km and 2km IFOs at LHO, known as H1 and H2,
respectively, are not suppressed by the ORF, which is identically
unity for colocated, coaligned IFOs.  Since H1 and L1 have
comparable sensitivities, the most significant factor in comparing
H1-H2 to L1-A1 sensitivity is the relative sensitivities of A1 and H2.
Since H2 was about a factor of 50 (in power) more sensitive than A1 during S4,
averaged across the band from 905\,Hz to 925\,Hz, we would expect an
H1-H2 correlation measurement during S4 to be a factor of 7 or better
more sensitive than L1-A1 as a measure of $\Omega_{\text{gw}}(f)$ at these
frequencies.  However, the fact that H1 and H2 share the same physical
environment at LHO necessitates a careful consideration of correlated
noise which is ongoing \cite{Fotopoulos:2006}.

Work is also currently underway to search for a SGWB at frequencies
around 900\,Hz by correlating data from the Virgo IFO with
the resonant bar detectors AURIGA, EXPLORER, and
NAUTILUS \cite{Cuoco:2006}.

Finally, an indirect limit can be set on SGWB strength due to the
energy density in the associated gravitational waves themselves, which
is given by
\begin{equation}
  \rho_{\text{gw}} = \rho_{\text{crit}}\int_{0}^\infty
  \frac{\Omega_{\text{gw}}(f)}{f}\,df
\end{equation}
The most stringent limit is on a cosmological SGWB, set by the success
of big-bang nucleosynthesis, is
$\rho_{\text{gw}}/\rho_{\text{crit}}\le 1.1\times
10^{-5}$ \cite{Maggiore:2000}. In comparison, a background of the
strength constrained by our measurement, $\Omega_R=\omegaLim$, would
contribute about {$2\times 10^{-2}$} to
$\rho_{\text{gw}}/\rho_{\text{crit}}$, if it were confined to the most
sensitive region between 905\,Hz and 925\,Hz.  (Spread over the full
range of integration $850\un{Hz}\le f\le 950\un{Hz}$, it would
contribute {$1\times 10^{-1}$}.)  Note, however, that this
nucleosynthesis bound is not relevant for a SGWB of astrophysical
origin.

\section{FUTURE PROSPECTS}
\label{s:future}

LIGO's S5 science run began in November 5 with the aim of collecting
one year of coincident data at LIGO design sensitivity.  ALLEGRO
has also been in operation over that time period, so the measurement
documented in this paper could be repeated with S5 data.  Such a
measurement would be more sensitive due to L1's roughly fivefold
reduction in strain noise power at 900\,Hz between S4 and S5, and
because of the larger volume of data (roughly 20 times as much).
Those two improvements could combine to lead to an improvement of
about an order of magnitude in $\Omega_R$ sensitivity.  However, no
immediate plans exist to carry out an analysis with S5 data, because
this incremental quantitative improvement in sensitivity would still
leave us far from the level needed to detect a cosmological background
consistent with the nucleosynthesis bound, or an astrophysical
background arising from a realistic model.

Additionally, the greater sensitivity of the H1-H2 pair means that a
background detectable with L1-A1 would first be seen in H1-H2.  In the
event that a ``surprise'' correlation is seen in H1-H2 which cannot be
attributed to noise, correlation measurements such as LLO-ALLEGRO and
Virgo-AURIGA could be useful for confirming or ruling out a
gravitational origin.

\section{CONCLUSIONS}

We have reported the results of the first truly heterogeneous
cross-correlation measurement to search for a stochastic
gravitational-wave background.  While the upper limit of $\strainLim$
on the strain of the SGWB corresponds to a hundredfold improvement
over the previous direct upper limit on $\Omega_{\text{gw}}(f)$ in
this frequency band \cite{Astone:1999}, the amplitude of conceivable
spectral shapes is already constrained more strongly by results at
other frequencies \cite{s3sgwb,s4sgwb}.  The lasting legacy of this work is
thus more likely the overcoming of technical challenges of
cross-correlating data streams from instruments with significantly
different characteristics.  Most obviously, we performed a coherent
analysis of data from resonant-mass and interferometric data, but
additionally the data were sampled at different rates, ALLEGRO data
were heterodyned and therefore complex in the time domain, and
entirely different methods were used for the calibrations of both
instruments.  Lessons learned from this analysis will be valuable not
only for possible collaborations between future generations of
detectors of different types, but also between interferometers
operated by different collaborations.

\begin{acknowledgments}
The authors gratefully acknowledge the support of the United States
National Science Foundation for the construction and operation of
the LIGO Laboratory and the Particle Physics and Astronomy Research
Council of the United Kingdom, the Max-Planck-Society and the State
of Niedersachsen/Germany for support of the construction and
operation of the GEO600 detector. The authors also gratefully
acknowledge the support of the research by these agencies and by the
Australian Research Council, the Natural Sciences and Engineering
Research Council of Canada, the Council of Scientific and Industrial
Research of India, the Department of Science and Technology of
India, the Spanish Ministerio de Educacion y Ciencia, The National
Aeronautics and Space Administration, the John Simon Guggenheim
Foundation, the Alexander von Humboldt Foundation, the Leverhulme
Trust, the David and Lucile Packard Foundation, the Research
Corporation, and the Alfred P. Sloan Foundation.
The ALLEGRO observatory is supported by the National Science
Foundation, grant PHY0215479.

This paper has been assigned LIGO Document Number LIGO-P050020-08-Z.
\end{acknowledgments}



\begin{thebibliography}{99}

\bibitem{Christensen:1992}
N.~Christensen, \prdoldref{46}{5250}{1992}

\bibitem{Allen:1997}
B.~Allen,
in {\it Proceedings of the Les Houches School on Astrophysical Sources of
Gravitational Waves, Les Houches, 1995},
edited by J.~A.~Marck and J.~P.~Lasota (Cambridge, 1996), p.~373.

\bibitem{Maggiore:2000}
M.~Maggiore,
\textit{Phys.~Rep.}\ \textbf{331}, 283 (2000).

\bibitem{Gasperini:1993}
  M.~Gasperini and G.~Veneziano \textit{Astropart.\ Phys.}\
  \textbf{1}, 317 (1993).

\bibitem{Gasperini:2003}
  M.~Gasperini and G.~Veneziano \textit{Phys.~Rep.}\ \textbf{373}, 1
  (2003).

\bibitem{Buonanno:1997}
  A.~Buonanno, M.~Maggiore, and C.~Ungarelli,
  \prdoldref{55}{3330}{1997}

\bibitem{Grishchuk:1975}
  L.~P.~Grishchuk, \textit{Sov.\ Phys.\ JETP} \textbf{40}, 409 (1975).

\bibitem{Grishchuk:1997}
  L.~P.~Grishchuk, \cqgref{14}{1445}{1997}

\bibitem{Starobinsky:1979}
  A.~A.~Starobinsky, \textit{Pis'ma Zh.\ Eksp.\ Teor.\ Fiz.}\
  \textbf{30}, 719 (1979).

\bibitem{Kosowsky:1992}
  A.~Kosowsky, M.~S.~Turner, and R.~Watkins, \prlref{69}{2026}{1992}

\bibitem{Apreda:2002}
  R.~Apreda, M.~Maggiore, A.~Nicolis, and A.~Riotto, \textit{Nucl.\
    Phys.\ B} \textbf{631}, 342 (2002).

\bibitem{Caldwell:1992}
  R.~R.~Caldwell and B.~Allen.  \prdoldref{45}{3447}{1992}

\bibitem{Damour:2000}
  T.~Damour and A.~Vilenkin, \prlref{85}{3761}{2000}

\bibitem{Damour:2005}
  T.~Damour and A.~Vilenkin, \prdref{71}{063510}{2005}

\bibitem{Regimbau:2001}
  T.~Regimbau and J.~A.~de~Freitas Pacheco,
  \textit{Astron.\ and Astrophys.}\ {\bf 376}, 381 (2001).

\bibitem{Regimbau:2006}
  T.~Regimbau and J.~A.~de~Freitas Pacheco,
  \textit{Astron.\ and Astrophys.}\ {\bf 447}, 1 (2006).

\bibitem{Coward:2002}
  D.~M.~Coward, R.~R.~Burman, and D.~G.~Blair,
  \textit{Mon.\ Not.\ R.\ Astron.\ Soc.}\ \textbf{329}, 411 (2002).

\bibitem{Cooray:2004}
  A.~Cooray,
  \textit{Mon.\ Not.\ R.\ Astron.\ Soc.}\ \textbf{354}, 25 (2004).

\bibitem{Allen:1999}
  B.~Allen and J.~D.~Romano, \prdref{59}{102001}{1999}

\bibitem{Astone:1999}
P.~Astone et al.,
\textit{Astronomy and Astrophysics} {\bf 351}, 811 (1999).

\bibitem{s4ALLEGROcal} 
M.~P.~McHugh and W.~W.~Johnson (2006), LIGO Report, \\
\href{http://www.ligo.caltech.edu/docs/T/T060096-00.pdf}
{\texttt{http://www.ligo.caltech.edu/docs/T/T060096-00.pdf}}

\bibitem{s1sgwb} B.~Abbott et al.\ (LIGO Scientific
  Collaboration), \prdref{69}{122004}{2004}

\bibitem{s3sgwb} B.~Abbott et al.\ (LIGO Scientific
  Collaboration), \prlref{95}{221101}{2005}

\bibitem{s4sgwb} B.~Abbott et al.\ (LIGO Scientific
  Collaboration), \xxx{astro-ph}{0608606}

\bibitem{Finn:2001} L.~S.~Finn and A.~Lazzarini,
  \prdref{64}{082002}{2001}

\bibitem{Flanagan:1993}
\'E.\'E.~Flanagan, \prdoldref{48}{2389}{1993}.

\bibitem{Whelan:2006}
J.~T.~Whelan, \cqgref{23}{1181}{2006}

\bibitem{Brustein:1995ah}
  R.~Brustein, M.~Gasperini, M.~Giovannini and G.~Veneziano,
  Phys.\ Lett.\ B {\bf 361}, 45 (1995)
  [arXiv:hep-th/9507017].

\bibitem{Bose:2005fm}
  S.~Bose,
  Phys.\ Rev.\ D {\bf 71}, 082001 (2005)
  [arXiv:astro-ph/0504048].

\bibitem{coward}
D.~M.~Coward, R.~R.~Burman, and D.~G.~Blair, 
Mon.~Not.~R.~Astron.~Soc. {\bf 324}, 1015 (2001).

\bibitem{ferrari}
V.~Ferrari, S.~Matarrese, and R.~Schneider, 
Mon.~Not.~R.~Astron.~Soc. {\bf 303}, 247 (1999).

\bibitem{McHugh:2005}
M.~P.~McHugh et al., \cqgref{22}{S965}{2005}

\bibitem{Morse:1997}
A.~Morse et al., \prdref{59}{062002}{1997}.

\bibitem{bias}
A.~Lazzarini (2004), LIGO Report, \\
\href{http://www.ligo.caltech.edu/docs/T/T040128-02.pdf}
{\texttt{http://www.ligo.caltech.edu/docs/T/T040128-02.pdf}}

\bibitem{ovlpwin}
A.~Lazzarini and J.~Romano (2004), LIGO Report, \\
\href{http://www.ligo.caltech.edu/docs/T/T040089-00.pdf}
{\texttt{http://www.ligo.caltech.edu/docs/T/T040089-00.pdf}}

\bibitem{Whelan:2005} J.~T.~Whelan et al,
\cqgref{22}{S1087}{2005}

\bibitem{Welch:1967}
  P.~D.~Welch,
  \textit{IEEE Transactions on Audio and Electroacoustics}
  \textbf{AU-15}, 70 (1967)

\bibitem{s1ligo}
B.~Abbott et al., \textit{Nucl.~Instrum.~Methods A}
\textbf{517}, 154 (2004).

\bibitem{s4cal}
A.~Dietz et al (2005), LIGO Report, \\
\href{http://www.ligo.caltech.edu/docs/T/T050262-00.pdf}
{\texttt{http://www.ligo.caltech.edu/docs/T/T050262-00.pdf}}

\bibitem{Bose:2003nb}
S.~Bose et al.,
\cqgref{20}{S677}{2003}

\bibitem{ALLEGRO} http://gravity.phys.lsu.edu/

\bibitem{Harry:1999}
Gregory M. Harry, thesis, University of Maryland 1999

\bibitem{Fotopoulos:2006}
  Nickolas V.~Fotopoulos for the LIGO Scientific Collaboration
  \cqgref{23}{S693}{2006}

\bibitem{Cuoco:2006} Elena Cuoco for the Virgo Collaboration, ``Virgo
  Data Analysis for C6 and C7 engineering runs'', presentation from
  the XIth Marcel Grossmann Meeting, July 2006.
  \href{http://wwwcascina.virgo.infn.it/vsb/slides/Cuoco_MG11.ppt}
  {\texttt{http://wwwcascina.virgo.infn.it/vsb/slides/Cuoco\_MG11.ppt}}
  To appear in the proceedings of the XIth Marcel Grossmann Meeting,
  July 2006.

\end{thebibliography}
\end{document}